\documentclass{aa}
\usepackage{graphicx}
\usepackage{txfonts}
\usepackage{mathtools}
\usepackage{soul}
\usepackage{hyperref}
\hypersetup{
  colorlinks,
  citecolor=blue,
  linkcolor=blue,
  urlcolor=blue}

\DeclarePairedDelimiter\abs{\lvert}{\rvert}
\DeclarePairedDelimiter\norm{\lVert}{\rVert}

\begin{document}

    \title{Scalable stellar evolution forecasting:}
    \subtitle{Deep learning emulation versus hierarchical nearest-neighbor interpolation}
    \titlerunning{Scalable stellar evolution forecasting}

    \author{
        K.~Maltsev\inst{1,2}\and
        F.~R.~N.~Schneider\inst{1,3}\and
        F.~K.~R{\"o}pke\inst{1,2} \and
        A.~I.~Jordan\inst{1} \and
        G.~A.~Qadir\inst{1}\and
        W.~E.~Kerzendorf\inst{4, 5} \and
        K.~Riedmiller\inst{1} \and
        P.~van~der~Smagt\inst{6,7}        
        }
    \institute{
        Heidelberger Institut f{\"u}r Theoretische Studien,
        Schloss-Wolfsbrunnenweg 35, D-69118 Heidelberg, Germany\\
        \email{kiril.maltsev@h-its.org}
        \and
        Zentrum f\"ur Astronomie der Universit\"at Heidelberg, Institut f\"ur Theoretische Astrophysik, Philosophenweg 12, D-69120 Heidelberg, Germany
        \and
        Zentrum f\"{u}r Astronomie der Universit\"{a}t Heidelberg, Astronomisches Rechen-Institut, M\"{o}nchhofstr. 12-14, D-69120 Heidelberg, Germany
        \and
        Department of Physics and Astronomy, Michigan State University, East Lansing, MI 48824, USA
        \and
        Department of Computational Mathematics, Science, and Engineering, Michigan State University, East Lansing, MI 48824, USA
        \and
        Machine Learning Research Lab, Volkswagen AG, Munich, Germany
        \and
        Faculty of Informatics, E\"{o}tv\"{o}s Loránd University, Budapest, Hungary
    }

    \date{Received 08 June 2023; Accepted 05 October 2023}
    
    \abstract{Many astrophysical applications require efficient yet reliable forecasts of stellar evolution tracks. One example is population synthesis, which generates forward predictions of models for comparison with observations. The majority of state-of-the-art rapid population synthesis methods are based on analytic fitting formulae to stellar evolution tracks that are computationally cheap to sample statistically over a continuous parameter range. The computational costs of running detailed stellar evolution codes, such as MESA, over wide and densely sampled parameter grids are prohibitive, while stellar-age based interpolation in-between sparsely sampled grid points leads to intolerably large systematic prediction errors. In this work, we provide two solutions for automated interpolation methods that offer satisfactory trade-off points between cost-efficiency and accuracy. We construct a timescale-adapted evolutionary coordinate and use it in a two-step interpolation scheme that traces the evolution of stars from zero age main sequence all the way to the end of core helium burning while covering a mass range from ${0.65}$ to $300 \, \mathrm{M_\odot}$. The feedforward neural network regression model (first solution) that we train to predict stellar surface variables can make millions of predictions, sufficiently accurate over the entire parameter space, within tens of seconds on a 4-core CPU. The hierarchical nearest-neighbor interpolation algorithm (second solution) that we hard-code to the same end achieves even higher predictive accuracy, the same algorithm remains applicable to all stellar variables evolved over time, but it is two orders of magnitude slower. Our methodological framework is demonstrated to work on the \textit{MESA Isochrones and Stellar Tracks} \citep{Choi2016} data set, but is independent of the input stellar catalog. Finally, we discuss the prospective applications of these methods and provide guidelines for generalizing them to higher dimensional parameter spaces.}
    

    \maketitle
%

\section{Introduction}\label{sec:introduction}
Several fields of astrophysics require fast and cost-efficient predictive models of stellar evolution for their deployment at scale. These include stellar population synthesis, $N$-body dynamics models of stellar clusters \citep[e.g.,][]{Kamlah}, iterative optimization-based stellar parameter estimation methods \citep[e.g.,][]{Bazot2012}, and large-scale galactic and cosmic evolution simulations \citep[e.g.,][]{Springel}  that require a stellar sub-grid physics. 

For example, the \textsc{Bonn Stellar Astrophysics Interface} 
\citep[BONNSAI;][]{Schneider2014} is a Bayesian framework that allows for the testing of stellar evolution models and (if the test is passed) to infer fundamental stellar model parameters given the observational data. Determination of fundamental stellar parameters that best match the observation requires costly iterative optimization procedures, such as Markov chain Monte Carlo nested sampling techniques, which need a large number of evaluations over a quasi-continuous parameter space for convergence to the best-fit model. In order to reduce systematic estimation errors, BONNSAI requires a stellar parameter grid to be as dense as possible.

However, there are costly computational demands arising from the traditional method of running a detailed stellar evolution code over a dense rectilinear grid in a stellar parameter space:  
for a fixed grid spacing, the number of stellar tracks to evolve scales to the power of the dimensionality of the fundamental stellar parameter space. The most important parameters of single star evolution are the: age, $\tau$; initial mass, $M_\mathrm{ini}$,  at the zero age main sequence (ZAMS); initial metallicity, $Z_\mathrm{ini}$; and initial rotation velocity, $v_\mathrm{ini}$. For stars of $M_\mathrm{ini} > 8 \, M_\odot$, the binary interaction effects become increasingly important: 71\% of all O-stars interact with a companion and for over half of them, this takes place during the main sequence evolution \citep{Sana2012}. Therefore, in order to evolve massive stars, the parameter space needs be expanded to cover eight dimensions $(\tau_1, M_\mathrm{ini, 1}, M_\mathrm{ini,2 }, v_\mathrm{ini,1}, v_\mathrm{ini,2}, Z_\mathrm{ini}, P_\mathrm{ini}, \epsilon)$ in general, where $P_\mathrm{ini}$ is the initial period, $\epsilon$ the eccentricity of the binary orbit, and $\tau_1 \simeq \tau_2$ to a good approximation.

\textsc{Modules for Experiment in Stellar Astrophysics}  
\citep[MESA;][]{Paxton2011} 
is an example of a detailed one-dimensional (1D) stellar evolution code with a modular structure, which allows us to update the adopted physics when generating stellar evolution tracks; for instance, the equation of state, the mass loss recipe, and the opacity tables. When evolving stars numerically over a wide and densely sampled parameter grid with MESA, there are two main computational challenges: 1) the computational cost associated with running the code over the large grid size and 2) the numerical instabilities. To overcome the latter, substantial manual effort is required to push a simulation past failure points by reconfiguring the code and by checking for unphysical results. The manual action mainly involves the adaptation of spatial mesh refinement and time step control strategy, as well as of the error tolerance thresholds in stellar model computation, to make sure the solvers converge over each evolutionary phase within a reasonable computation time.

The problem of prohibitive computational costs has been addressed in three different ways: 1) the stellar evolution tracks have been approximated by analytic fitting formulae; 2) the output of detailed stellar evolution codes over a discrete parameter grid has been interpolated; and 3) cost-efficient surrogate models of stellar evolution have been constructed. Below, we summarize these main approaches.

 The \textsc{single star evolution} (SSE) package \citep{Hurley2000} consists of analytic stellar evolution track formulae predicting stellar luminosity, radius and core mass as functions of the age, mass and metallicity of the star. Separate formulae are applied to each evolutionary phase and the duration of each phase is estimated from physical conditions. Along with analytical expressions from stellar evolution theory, the SSE package was obtained by fitting polynomials to the set of stellar tracks by \cite{Pols1998}. The fitting formulae method has been extended to predict the evolution of binary systems by including analytical prescriptions for mass transfer, mass accretion, common-envelope evolution, collisions, supernova kicks, angular momentum loss mechanisms and tides \citep{Hurley2002}. At present, the fitting formulae are often used in connection with rapid binary population synthesis codes, for example \textsc{Compact Object Mergers: Population Astrophysics \& Statistics} \citep[COMPAS;][]{Riley_2022}, and stellar $N$-body dynamics codes. However, there are two main drawbacks: 1) the fixed (rather than modular) input physics and 2) the limited set of predicted output variables, which (depending on the astrophysical application) may be not all the variables of interest. 
A re-derivation of analytic fitting formulae for a new set of stellar tracks is non-trivial \citep{church_tout_hurley_2009, Tanikawa}. Overall, the analytic approach is not sustainable, since it would need to be reiterated after each update in stellar input physics.

The interpolation of tracks pre-computed by a detailed code is an alternative to analytic fitting. \cite{BrottB} interpolate stellar variables in a $(M_\mathrm{ini}, v_\mathrm{ini}, \tau)$ parameter space. 
For each stellar age, the two nearest neighbors (from above and from below) in initial mass are selected first, and then, for each of the two initial masses, the two nearest neighbors in initial rotational velocity are chosen. The values of stellar evolution variables, at each stellar age, are computed from these four neighboring grid points by a sequence of linear interpolations in the sampled parameter space.
The scope of the interpolation method is restricted to the main sequence evolution of stars. Instead of the stellar age, the fractional main sequence lifetime is used as interpolation variable.

Following a different approach to interpolation of stellar tracks, the \textsc{Method of Interpolation for Single Star Evolution} code \citep[METISSE;][]{Agrawal} takes as its input a discrete single-star parameter grid and uses interpolation by a piece-wise cubic function to generate new stellar tracks in-between the sampled initial mass grid points at fixed metallicity. The parameter space covers the initial mass range from 0.5 to $50~\mathrm{M_\odot}$, and stars are evolved up to the late stages beyond core helium burning. Instead of stellar age, the interpolation scheme uses a uniform basis known as \textsc{Equivalent Evolutionary Points} \citep[EEP;][]{Dotter2016} to model the evolutionary tracks. The EEP coordinate quantifies the evolutionary stage of a star based on physical conditions, derived from numerical values of evolutionary variables (e.g., depletion of central hydrogen mass fraction to a threshold value), which are readily identifiable for different evolutionary tracks. For any given stellar age, an isochrone is constructed by identifying which EEP coordinate values are valid for that age as function of $M_\mathrm{ini}$. For each fixed EEP value, an ordered $M_\mathrm{ini}-\tau$ relation is constructed over the available grid points and interpolated over. In a second step, $M_\mathrm{ini}$ is used as independent variable to obtain stellar properties by another round of interpolation.
Reliable and fast stellar track interpolation with the EEP method has originally been demonstrated upon \textsc{MESA Isochrones and Stellar Tracks} \citep[MIST;][]{Choi2016}, a catalog of stellar evolution tracks over a grid space covering the age, initial mass and initial metallicity parameters.
METISSE is a more general alternative to SSE, because it may take any single star grid (at fixed initial metallicity), produced as output of a detailed stellar evolution code, as its input; namely, it is not tied to specific input physics adopted to generate the stellar tracks. 

Apart from METISSE, there are the \textsc{combine} \citep{kruckow}, \textsc{sevn} \citep[][in its latest version]{iorio}, and  \textsc{posydon} \citep{fragos} population synthesis codes that interpolate grids of detailed single or binary evolution simulations. Interpolation in \textsc{combine} is based on the method of \citet{BrottB} while in \textsc{sevn}, single star evolution is divided into sub-phases analogous to the EEP method, and interpolation is performed over each sub-phase using a fractional time coordinate relative to duration of each sub-phase. Evolution of the binary companion and interaction effects are approximated using analytic fitting formulae. Since the procedure to construct the uniform EEP basis cannot be trivially automatized, the pre-processing steps to identify EEPs, to define appropriate interpolation functions and also to down-sample the stellar evolution catalog to reduce memory costs need to be re-iterated after each stellar grid update \citep[see, e.g.,\ the \textsc{TrackCruncher} pre-processing modules,][in the context of \textsc{sevn}]{iorio}.\\
In contrast, \textsc{posydon} interpolates output of detailed binary evolution simulations with MESA. The EEP-based interpolation method is not directly applicable to binary evolution tracks, because EEPs must be strictly ordered a priori while binary interaction, which can set on at any time, may change their order. Therefore, in \textsc{posydon} interpolation needs to be preceded by classification of binary evolution phase and separate interpolation schemes are to be applied over each of them.\\   
Finally, the third way is to build a prediction-making tool that allows for  the replacement of the output of cost-intensive detailed up-to-date stellar evolution code such MESA with a cost-efficient imitation model (emulator or surrogate) of the original. Emulation, or surrogate modeling, is a pragmatic but reliable reproduction of the output generated by an expensive computer experiment. The predictive surrogate model is constructed by training a supervised machine learning (ML) algorithm on a stellar evolution tracks data base pre-computed with the original code over a discrete parameter grid. 
A well-trained model will not only efficiently reproduce stellar tracks at the parameter grid points it has seen during training, but it will be capable of generating accurate predictions of tracks in-between the grid points, thanks to the capability to generalize it acquired by training. Once constructed, the emulator can be used as a package to generate predictions of stellar variables of interest, instead of running the original detailed stellar evolution code such as MESA over a quasi-continuous parameter range or storing the catalog data in computer memory for interpolation. 
Using the emulator package saves energy costs, speeds up generation of output predictions over a dense grid by several orders of magnitude and reduces human effort of running models. The speed-up is owed to the efficiency of input-to-output mapping by machine learning algorithms. The disadvantage 
is the introduction of prediction errors by the trained model, which reproduces stellar tracks with a finite precision. Therefore, when training machine learning models, the main task is to achieve reliable generalization over the parameter space with a prediction inaccuracy of stellar variables of interest that is tolerable for inference and astrophysical applications. 

Surrogate modeling of stellar evolution has yet not been explored extensively at widths of the parameter range necessary for more general applicability. \citet{Li2022} used a Gaussian process regression (GPR) to emulate stellar tracks in a five-dimensional (5D) parameter space, but the initial mass range covered by the predictive models is restricted to the solar-mass neighborhood $M_\mathrm{ini} \in (0.8, 1.2) \, \mathrm{M_\odot}$ and to evolutionary sequences from the Hayashi line onward through the main sequence up to the base of the red giant branch. 
Also, GPR-based emulators have been used, for example, for parameter space exploration of state-of-art rapid binary population synthesis codes such as COMPAS \citep{Barrett2017, Taylor2018}. Due to the data set size limitation for the applicability of GPR, it is not the ideal tool for emulating a large stellar model grid. Thus, we seek for other ML based models instead. The feedforward neural network algorithm proved itself as promising in previous surrogate modeling works, for instance: \citet{Scutt} emulated 25 stellar output variables (classic photometric variables, asteroseismic quantities, and radial and dipole mode frequencies) over a $(M_\mathrm{ini}, Z_\mathrm{ini})$ grid space of stars in or near the $\delta$ Scuti instability strip using neural networks, along with a principal component analysis to reduce the output dimension to nine.
\citet{Lyttle} emulated five variables of red dwarfs, sun-like stars, and subgiants in a 5D input parameter space. While these  are high-dimensional problems that have been successfully addressed by neural networks, aspects that the problem settings have in common include: the mass range considered is relatively narrow, $M_\mathrm{ini} \in (1.3, 2.2) \, \mathrm{M_\odot}$ and  $M_\mathrm{ini} \in (0.8, 1.2) \, \mathrm{M_\odot}$, respectively; and the evolutionary sequences cover the pre-main sequence and only part of the main sequence, or main sequence and subgiant phase, respectively. More widely in context of stellar astrophysics, supervised machine learning has been applied  to solve the inverse problem of mapping observables to models. For example, a variant of the random forest regression model \citep{Bellinger2016} and invertible neural networks \citep{Ksoll} have been trained to predict fundamental stellar parameters in a high-dimensional parameter space given a set of observational variables.  Again, however, the predictive models were restricted to an initial mass range and evolutionary sequences of stars narrower (e.g., main sequence evolution of $M_\mathrm{ini} \in (0.7, 1.6) \, \mathrm{M_\odot}$ stars in \citealt{Bellinger2016}) than those presented in this work, where we consider an initial mass range from red dwarfs to very massive stars evolved from the ZAMS up to the end of core helium burning. 

In this work, we provide two proof-of-concept solutions of automated single star interpolation schemes over a wide parameter span, which (in contrast to the EEP-based interpolation method) do not require mapping out points of interest in stellar parameter space; this is because they are constructed based on a timescale-adapted evolutionary coordinate that we introduce, whose computation can be easily automated. Using the latter for constructing more general interpolation models has a potential applicability to larger parameter spaces, such as those found in stellar binaries. The first solution we develop is a surrogate model of stellar evolution, constructed with supervised machine learning. The second is a stellar-catalog-based hierarchical nearest-neighbor interpolation (HNNI) method. These feature two different trade-off points between efficiency and accuracy of predictions: depending on astrophysical application, either the one or the other is preferable.

This paper is organized as follows. In Sect.~\ref{sec:methods}, we describe the methods common to both interpolation scheme solutions that we have developed: the regression problem that is addressed, the data base used for constructing predictive models, the timescale-adapted evolutionary coordinate (which is used as the primary interpolation variable), and the performance scores that assess the quality of the predictions. 
In Sect.~\ref{sec:ffNN_GPR}, we outline how the two interpolation scheme solutions are set up. For the surrogate model, we report on the choice of loss function, the selection of machine learning model class, and its hyperparameter optimization. For the interpolation-based solution, we explain how HNNI works and how it differs from interpolation models from previous work.  
In Sect.~\ref{sec:results}, we present our results, obtained with both the supervised machine learning and the HNNI. The paper is concluded in Sect.~\ref{sec:concl} with a summary of results, limitations, and an outlook on possible future developments. 

\section{Methods}
\label{sec:methods}
In Sect.~\ref{sec:regr}, we define the problem which is addressed by two different predictive frameworks (surrogate modeling of stellar evolution and catalog-based hierarchical nearest neighbor interpolation) and we motivate the two-step approach to fitting stellar evolution tracks. In Sect.~\ref{sec:s}, the timescale-adapted evolutionary coordinate is introduced, which we used to set up reliable predictive frameworks in the two-step interpolation scheme. In Sect.~\ref{sec:dat}, the methods to prepare the data base are described: a nonlinear sampling density segmentation of the initial mass parameter space and a data augmentation routine for the core helium burning phase. This data base is used as catalog for interpolation of tracks by HNNI and as training data for constructing surrogate models. Finally,  Section~\ref{sec:perf} outlines how we evaluate predictive performance of our models based on error metrics.        
 
\subsection{Regression problem formulation}
\label{sec:regr}
In 1D stellar evolution codes such as MESA, stellar evolution is modeled as a deterministic initial value problem and observables are predicted by cost-intensive numerical time integration of differential equations. Instead, we formulated the prediction of observables as a regression problem, which is to be addressed by supervised machine learning or by catalog-based interpolation. In a regression problem, the goal is to predict output target variables from input regressor variables. But in the surrogate modeling case,  the data-driven approach is used to learn the mapping, instead of programming the rules that map the input to the output.
We constrained the problem to predicting three stellar surface observables, namely, log-scaled luminosity, $Y_L = \log L/\mathrm{L_\odot}$; effective temperature, $Y_T = \log T_\mathrm{eff}/ \mathrm{K}$; and surface gravity, $Y_g = \log g/[\mathrm{cm} \cdot \mathrm{s}^{-2}]$. These are the target variables to be predicted for a given input of age, $\tau,$ and initial mass, $M_\mathrm{ini}$, of an isolated non-rotating single star, at a fixed solar-like initial metallicity $Z_\mathrm{ini}=Z_\odot$.

Stars evolve on different timescales, depending on the evolutionary phase they undergo, on their masses as well as on other stellar parameters. Therefore, stellar track fitting across different evolutionary phases and initial masses is a temporal multiscale problem. We confirm the conclusion of \cite{Li2022}, namely, that the naive approach of training a machine learning surrogate model $f_\mathrm{ML} \colon (\tau, M_\mathrm{ini}) \mapsto Y$ to predict the observable $Y$, by operating directly on (scaled) age, $\tau$, does not result in accurate enough predictions of the post-main sequence evolution (see Fig.~\ref{fig:naive-twostep} for an illustration). Instead, we set up a two-step interpolation scheme:
\begin{align*}
&\text{Step 1 (age proxy fit)}\ f_1 \colon (\log \tau, \log M_\mathrm{ini}) \mapsto s, \\
&\text{Step 2 (observables fit)}\ f_2 \colon (s, \log M_\mathrm{ini}) \mapsto (Y_L, Y_T, Y_g).
\end{align*}
Here, the evolution of stellar surface variables is modeled as function of a timescale-adapted evolutionary coordinate $s$ (an age proxy) instead of the age, $\tau$ (step 2).
The transition from stellar age to the age proxy is accomplished by a second predictive model (step 1).

We find that the fits  of the post-main sequence evolutionary stages resulting from this two-step interpolation scheme are orders of magnitude more accurate, as assessed by standard statistical performance scores, than the direct naive fit. We take the logarithm 
of initial mass values, in order to exploit the approximate mass-luminosity power law relation, which is a linear variable dependence in log-log space.

\subsection{The timescale-adapted evolutionary coordinate}
\label{sec:s}
The method of using a timescale-adapted evolutionary coordinate, or age proxy, instead of the age variable for fitting stellar evolution tracks  has been explored before in stellar astrophysics \citep[e.g.,][]{Jorgensen2005, Li2022}. The motivation for this re-parametrization is to reduce timescale variability. Stellar age at computation step, $i$,
\begin{equation*}
    \tau_i = \sum_{j=1}^i \delta t_j,
\end{equation*}
 is a monotonically increasing function which grows cumulatively at an adaptive step size, $\delta t_j$, after each step $j=1, \dots, i$ of numerical time integration of the differential equations describing stellar structure and evolution. The age proxy variable, 
\begin{equation*}
    s_i = \sum_{j=1}^i \delta s_j,
\end{equation*}
is constructed analogously, but here $\delta s_j$ is the increment in the star's Euclidean displacement in a diagram spanned by a set of its physical variables, obtained after the numerical time integration step $j=1, \dots, i$. For a parametric form of $\delta s$, \cite{Jorgensen2005} used the ansatz
\begin{equation*}
    \delta s_j = \sqrt{ \left| \Delta_{j,j-1}\log \frac{L}{\mathrm{L_\odot}} \right| ^2+ \left| \Delta_{j,j-1} \log \frac{T_\mathrm{eff}}{\mathrm{K}} \right|^2},
\end{equation*}
where $\Delta_\mathrm{j,j-1} X = X_j - X_\mathrm{j-1}$. By construction, this age proxy measures the increase in Euclidean path length of a star along its evolutionary track in the Hertzsprung-Russell (HR) diagram. More recently, \cite{Li2022} suggested another prescription
\begin{equation*}
    \delta s_j = \left(  \left| \Delta_\mathrm{j,j-1} \log \frac{g}{[\mathrm{cm} \, \mathrm{s}^{-2}]} \right|^2+\left| \Delta_\mathrm{j,j-1} \log \frac{T_\mathrm{eff}}{\mathrm{K}} \right|^2 \right)^c,
\end{equation*}
which they tailored to their problem formulation and parameter range. Their age proxy measures the displacement of the star in the Kiel diagram to the power of a parameter, $c$. After experimentation, they conclude that $c=0.18$ yields the most uniform distribution of the data they trained their models on. At the same time, the authors report fit inaccuracies at transition regions between consecutive evolutionary phases and over the fast ascension of the red giant branch. Over these phases (in contrast to the MS evolution) target variables change rapidly in time and vary unsteadily even as function of the age proxy. To cure this problem, we re-defined the timescale-adapted evolutionary coordinate by an altered prescription, whose effect is to not only smooth out transitions in-between stellar phases, but, additionally, to resolve the CHeB phase in a way that allows for reliable stellar track fitting; this is done by keeping the resolution of variability on the same numerical age proxy scale as the previous two phases.  To get there, we found a promising approach in returning to the original formulation by \cite{Jorgensen2005}, but extending it by a third variable that spans another dimension of the diagram, in which the Euclidean path length is calculated: 

\begin{equation*}
    \delta \tilde{s}_j = \sqrt{ \left| \Delta_\mathrm{j,j-1}\log \frac{L}{\mathrm{L_\odot}} \right|^2+ \left| \Delta_\mathrm{j,j-1} \log \frac{T_\mathrm{eff}}{\mathrm{K}} \right|^2+ \left| \Delta_\mathrm{j,j-1} \log \frac{\rho_{c}}{[\mathrm{g} \cdot \mathrm{cm}^{-3}]} \right|^2 }.
\end{equation*}

The motivation for introducing another variable into the computational prescription of the path length stems from the fact that during the stable CHeB, stars hardly displace in the HR diagram, although their nuclear composition and hydrodynamic properties undergo substantial changes. In order to adjust the path length prescription, we therefore sought for a suitable stellar-core-related variable. After experimental tests, we found that adding the log-scaled core density  $\log \rho_{c} / [\mathrm{g} \cdot\mathrm{cm}^{-3}]$ has the desirable effect of casting the variability of all target variables of interest onto a unified numerical scale across the three consecutive phases MS, RGB, CHeB, and across the wide initial mass range that we work with.\footnote{This age proxy computation prescription has the aforementioned desirable effects not only during these phase, but also during the pre-MS and post-CHeB phases, as shown in Fig.~\ref{fig:age-proxy2} in the appendix. Our age proxy construction therefore is a promising general candidate solution to the multiscale problem of stellar evolutionary track fitting beyond the evolutionary sequences considered in this work. It resolves prominent features (e.g.,\ the Henyey MS hook, MS turnoff, the Hertzsprung gap, the base and tip of the RGB, dredge-ups, helium flashes, blue loops, thermal pulsations on the asymptotic giant branch, and white dwarf cooling) across all evolutionary phases we tested over the wide initial mass span.} 

\begin{figure*}
      \centering
      \includegraphics[width=0.32\textwidth]{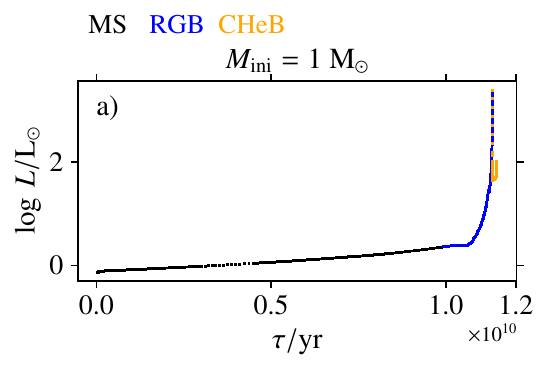}
      \includegraphics[width=0.32\textwidth]{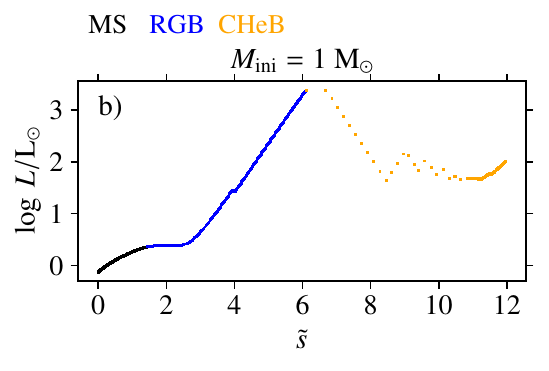}
      \includegraphics[width=0.32\textwidth]{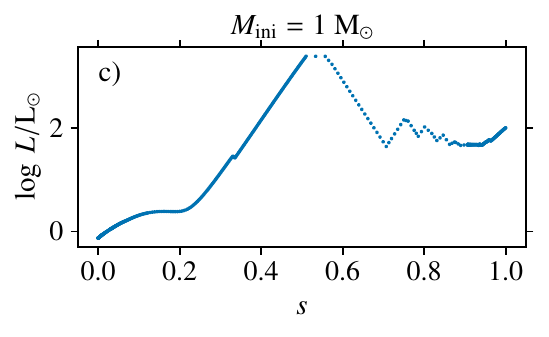}
      \caption{Luminosity series of a Sun-like star from the ZAMS up to TACHeB parametrized as function of stellar age, $\tau,$ (a) versus of the timescale-adapted evolutionary coordinate, $\tilde{s}$, before (b) versus after (c) CHeB data augmentation and normalization to $s$. The original MIST data contains phase labels for each model, which the predictive models (the surrogate model and HNNI) do not see.} 
    \label{fig:age-proxy}
\end{figure*}

We normalize the age proxy of each initial mass  to the range (0, 1). The star is on the ZAMS when $s=0$, while $s=1,$ when the star has reached the end of core helium burning.\footnote{The end of core helium burning is determined by the condition $X_\mathrm{He, central} \leq 10^{-3}$, where $X_\mathrm{He, central}$ is the central helium mass fraction.} 

\subsection{Data base}
\label{sec:dat}
\paragraph{Stellar evolution catalog:}
Here, we use MIST \citep{Choi2016} as an example data set upon which we formulate and demonstrate our method, as well as train and validate our predictive models. However, the method we develop is general and not specific to the MIST data set. We restrict the scope of the ages of stars to the evolutionary sequence from ZAMS to the terminal age of core helium burning (TACHeB), which is expected to account for $\simeq$ 99 \% of stellar observations (excluding compact object sequences). The initial mass parameter range, from 0.65 to 300 $\mathrm{M_\odot}$, is chosen as the entire initial mass span available in the MIST data set, over which stars are evolved through all three consecutive phases main sequence (MS), red giant branch (RGB), and core helium burning (CHeB). 
The wide initial mass range and, at the same time, the inclusion of the red giant as well as core helium burning phases have not been explored in previous work of stellar evolution surrogate modeling. We acknowledge that the 2D input parameter space is small compared to the size of the eight-dimensional (8D) parameter space required for general cost-efficient binary star modeling. We see our work as a first step toward a large-scale enterprise of stellar evolution surrogate modeling and of hierarchical interpolation in high-dimensional parameter space over wide parameter ranges, however, as a layout of basic methodology toward this end.

\paragraph{CHeB data augmentation:}
The MIST data set is generated with the MESA code, which by default outputs more stellar evolution models than what is included in the MIST data set for each $M_\mathrm{ini}$-dependent track. The number of models per track is ${\sim} 500$, with ${\sim} 250$ models on the MS, ${\sim} 150$ on the RGB before ignition of helium burning in the core, and  ${\sim} 100$ for the CHeB phase. While the MIST data set includes phase labels for each stellar model, the predictive models that we build are not exposed to this information. All the input information they are exposed to is the value of the age (proxy) and of initial mass of the star. While in the MIST data set, the CHeB phase is the least sampled among these three, it is the phase most difficult to fit. In particular, the helium flashes of low-mass stars, blue loops of upper main-sequence stars, and fast timescale dynamics of Wolf-Rayet stars during CHeB pose a challenge to fitting. To increase weight and accuracy of interpolation fits during the CHeB phase, we use local nearest-neighbor 1D linear interpolation of the training data (not of the test data) along the age proxy axis (for the step 2 fit) or along the scaled age axis (for the step 1 fit) during this phase. 
The net effect is an artificial increase in the CHeB training data by insertion of a sample in-between each pair of age proxy neighbors. Despite simplicity of this methodological step, we find the predictive performance of our best-fit models to be boosted by around half an order of magnitude decline in the mean squared error over the validation data (to which CHeB data augmentation is not applied) after switching on CHeB data augmentation of the training data. 
In Fig.~\ref{fig:age-proxy} the data pre-processing consisting of age proxy re-parametrization, normalization, and CHeB data augmentation is illustrated based on the example of the Sun-like stellar model.

\paragraph{Parameter space grid sampling:} 
A recommended standard routine for a homogeneous sampling of the parameter space that produces the data for training surrogate models is \textsc{Latin Hypercube Sampling} (LHS) \citep{McKay1979}. This is an efficient alternative to random uniform and rectilinear sampling methods for achieving homogeneity. Random sampling introduces sampling voids by consequence of statistical random clumping effects, while dense rectilinear sampling is too expensive in many problem settings. However, since the stellar evolution dependence on the initial mass parameter is strongly non-linear, a homogeneous population of parameter space is not the optimal sampling scheme. We work with the pre-computed MIST data set for which a segmented parameter sampling density across the initial mass range has already been pre-determined by the makers of the catalog, based on physics-informed considerations.   

\begin{figure}
  \centering
  \includegraphics[width=0.45\textwidth]{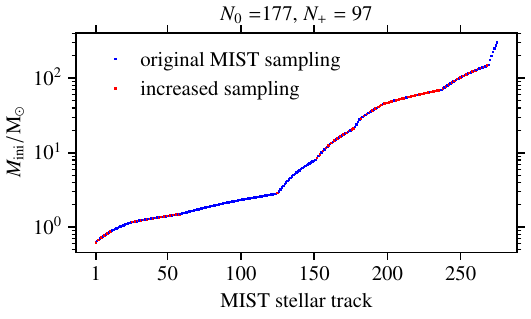}
  \caption{Original initial mass sampling in the MIST catalog (in blue) and the locally increased sampling (in red) that we used for training the surrogate models. The stock MIST catalog contains 177 solar metallicity stellar evolution tracks within the initial mass range $(0.65, 300) \, \mathrm{M_{\odot}}$. For our purposes, we expanded it to 274 to achieve the desired quality of predictive accuracy necessary for a general-purpose stellar evolution emulator.}
\label{fig:mist-sampling}
\end{figure}

\begin{table}[h!]
  \begin{center}
  \caption{Summary of the initial mass sampling density segmentation before ($\delta_0 M_\mathrm{ini}$) and after ($\delta M_\mathrm{ini}$) expanding the stock MIST data set.}
    \begin{tabular}{l l l c}        
    \hline\hline  
      
      Range ($M_\mathrm{ini}/ \mathrm{M_\odot}$) & $\delta_0 M_\mathrm{ini}$ [$\mathrm{M_\odot}$] & $\delta M_\mathrm{ini}$ [$\mathrm{M_\odot}$]  & $N_{M_\mathrm{ini}}$\\
      \hline
      $(0.65, 0.9)$  & 0.05  &  0.025 & 10\\
      $(0.9, 1.16)$ & 0.02 & 0.02 & 13\\
      $(1.16, 1.5)$ & 0.02 & 0.01  & 34\\
      $(1.5, 2.8)$ & 0.02 & 0.02 & 65  \\
      $(2.8, 3)$ & 0.2 & 0.1 & 2 \\
      $(3, 8)$ & 0.2 & 0.2 & 25 \\
      $(8, 21)$ & 1 & 0.5 & 26 \\
      $(21, 22)$ & 2 & 1 & 1\\
      $(22, 28)$ & 2 & 2 & 4\\
      $(28, 40)$ & 2 & 1 & 12 \\
      $(40, 45)$ & 5 & 1.25 & 4 \\
      $(45, 70)$ & 5 & 0.625 & 40  \\
      $(70, 150)$ & 5 & 2.5 & 32  \\
      $(150, 300)$ & 25 & 25 & 6 \\
      \hline
    \end{tabular}
  \label{Tab:table1}
  \end{center}  
\end{table}

In order to reach a high accuracy level of stellar track forecasts that is necessary for a general-purpose stellar evolution emulator across the entire initial mass range, we found that we need to locally increase the initial mass sampling. In practice, we increased the initial mass sampling in those parameter space sub-regions, where the fit quality was found to be worst, while we kept the MIST stock sampling intact, where the local fit accuracy was found to be satisfactory (see Fig.~\ref{fig:mist-sampling} and Table~\ref{Tab:table1} for a summary). For generating the additional stellar tracks, we used the MIST Web Interpolator\footnote{\url{https://waps.cfa.harvard.edu/MIST/interp_tracks.html}}, which works by applying the EEP-based method referred to in Sect.~\ref{sec:introduction}.
Our finding is that the final sampling required to reach the predictive accuracy goal varies substantially, depending on sub-region of parameter space: a least $\delta M_\mathrm{ini}/\mathrm{M_\odot}=0.01$ between $M_\mathrm{ini}/ \mathrm{M_\odot} \in (1.16, 1.5)$ and a largest $\delta M_\mathrm{ini}/\mathrm{M_\odot}=25$ between $M_\mathrm{ini}/\mathrm{M_\odot} \in (150, 300)$. 
 For the $M_\mathrm{ini}/\mathrm{M_\odot} \in (0.65, 0.9)$ interval, we double the sampling to correct for a systematic under-representation of red dwarfs in the stock MIST catalog as compared to the adjacent initial mass intervals. For the $M_\mathrm{ini}/\mathrm{M_\odot} \in (1.16, 1.5)$ interval, we double the sampling rate mainly because of complexity of shape changes in HR diagrams due to the helium flashes. In the interval $M_\mathrm{ini}/\mathrm{M_\odot} \in (1.5, 40)$, we hardly increase the sampling, except at transitions in-between neighbouring sampling segments at different rates, in order to smooth out transitions. The biggest increase in this range is within the interval $ M_\mathrm{ini}/\mathrm{M_\odot} \in (8, 21)$. 
We stress that our densest sampling region (the solar neighborhood initial mass range) is the same as in \cite{Li2022}, in \cite{Bellinger2016} and in \cite{Lyttle}, while at the same time our surrogate models evolve the stars further, up to end of CHeB, and cover a much wider initial mass range. \cite{Scutt} adopt a sampling of $\delta M_\mathrm{ini} = 0.02 \, \mathrm{M_\odot}$ over the range $M_\mathrm{ini}/\mathrm{M_\odot} \in (1.3, 2.2)$, comparable to ours. 
\\ At the high-mass end, the relative increase in sampling is greatest within the interval $M_\mathrm{ini}/\mathrm{M_\odot} \in (40, 70)$, where the increment step size $\delta M_\mathrm{ini}/\mathrm{M_\odot}$ was augmented from $5$ to $0.625$. We suspect that numerical challenges are the reason for unexpectedly sharp, peculiarly shaped changes in HR diagrams. Nevertheless, for the proof-of-concept, we assume that MIST offers a perfect data set, even when we suspect that it may be not. 

Naturally, the denser the grid sampling, the more accurate are the forecasts of surrogate models. We stress that depending on minimal performance benchmarks (as quantified by error scores) of a specific astrophysical application, the initial mass sampling required to reach that benchmark can be significantly sparser.

With the initial-mass parameter space sampling as described above, the total size $N_\mathrm{tot}$ of the data set amounts to $139016$. Shuffling it, we do a uniform random split of the $N_\mathrm{tot}$ into 85 \% training ($N_\mathrm{train}$) and 15 \% validation ($N_\mathrm{val}$) data sets. To the $N_\mathrm{train}$ data, we applied a CHeB data augmentation, which yields additional $N_\mathrm{aug}=32143$ samples, such that the expanded training data set is of size $N_\mathrm{train}'=N_\mathrm{train}+N_\mathrm{aug}$. This is the final data set on which we train different classes of surrogate models (for the first solution) or which we use as the catalog for interpolation (for the second solution).
         
\subsection{Performance evaluation}
\label{sec:perf}

\paragraph{Validation and test data:}
We used two schemes to evaluate performance of predictive models: the first (model validation) based on the validation data set, and the second (model testing) based on the test data set. The validation data consists of randomly selected grid points over the input domain (initial masses and evolutionary phases of stars). It is representative, since it has similar statistical properties as the training data. In contrast, for model testing, we aim to assess the trained model's capability to predict entire stellar tracks from ZAMS up to TACHeB for initial masses unseen during training. We choose this method of model testing since it is of main interest to obtain a predictive model that is capable of accurate interpolation over the space of fundamental stellar parameters. Only then can the traditional method of running expensive simulations over densely sampled grids be replaced by a surrogate model capable of sufficiently accurate generalization. As test data, we prepared another set of stellar tracks at 16 initial mass grid points, $\{M_\mathrm{ini}^\mathrm{test}/\mathrm{M_\odot}\}=\{$0.91, 1.51, 2.41, 4.1, 8.25, 16.25, 21.5, 31.5, 41, 51, 61, 83.75, 103.75, 155, 262.5, 295$\}$, which we held back from training. These were chosen at half of the grid step in the respective region of parameter space. This choice is motivated by the aim to test predictive accuracy at parameter space points that are farthest away from training grid points, where we likely probe the worst cases of complete stellar track predictions.\footnote{An alternative approach to choosing test initial mass grid points is to sample the initial mass range randomly, in order to obtain the statistically likely distribution of prediction errors of stellar tracks. Since we quantitatively probe statistical error distribution already on the validation data set by appropriate statistical error scores, we opt for the half-grid step approach to probe the worst cases instead.} 

\paragraph{Performance scores:}
A crucial ingredient for the optimization procedure of an automated interpolation method is a set of appropriately designed scores that quantify performance in a physically meaningful and numerically appropriate manner. Only with an adequately defined quantitative performance scoring, the automated interpolation scheme can be scaled up to higher dimensional fundamental stellar parameter spaces, which become too large for visual inspection based performance evaluation for comparing the predicted against the held-back test tracks.

For model validation on the validation data set, we look at residuals for each observable independently and at measures of overall predictive performance. A residual is a signed prediction error, $\epsilon_i = Y_i - \hat{Y}_i$, of a given prediction-label pair $(\hat{Y}_i, Y_i)$, and we evaluate it for each of the surface variables. We consider the following error scores that retain the physical significance of residuals: the mean residual $\overline{\epsilon} = (\epsilon_1 + \dots + \epsilon_{N_\mathrm{val}}) / N_\mathrm{val}$, the  most extremal under-prediction $\epsilon^+ = \max_{i = 1, \dots, N_\mathrm{val}} \{\epsilon_i\}$, and the most extremal over-prediction $\epsilon^- = \min_{i = 1, \dots, N_\mathrm{val}} \{\epsilon_i\}$. If $\epsilon^+$ and $\epsilon^-$ are close enough to zero over the entire set of validation data grid points, then there is no need to further stratify the performance evaluation.\footnote{The statistical performance assessment can be further stratified by applying the scoring prescriptions not globally over the entire initial mass range and over the full evolutionary sequence, but to confined sub-regions of parameter space. 
For instance, $\epsilon_L^+(1.2, 4.5)_\mathrm{RGB}$ is the most extremal under-prediction of log-scaled luminosity during the RGB phase within the initial mass segment $M_\mathrm{ini}/\mathrm{M_\odot} \in (1.2, 4.5)$.} Additionally, we use the following error scores to quantify overall predictive performance across the three surface variables: the Mean Squared Error (MSE) and the Mean Absolute Error (MAE). These scores are calculated from the squared residuals and from the absolute residuals, respectively, by taking the average variable by variable and over the three surface variables. We choose the MSE and the MAE, because these are standard choices for evaluating point forecasts generated by statistical learning models, but the physical significance is largely lost by averaging across surface variables. 
    
For model testing on the held-back test tracks at the 16 $M_\mathrm{ini}$ grid points stated above, we define and use the following error scores based on HR and Kiel diagrams:  $L2_\mathrm{HR}^+$ and $L2_\mathrm{K}^+$. For a single track in HR or in Kiel diagram at a particular initial mass, $M_\mathrm{ini}$, the $L2$ score measures the cumulative deviation between predicted track and held-back track,
\begin{equation*}
    L2(M_\mathrm{ini}) = \frac{1}{N(M_\mathrm{ini})}  \sum_{i=1}^{N(M_\mathrm{ini})} \norm*{\vec{v}_i - \hat{\vec{v}}_i}^2,
\end{equation*}
computed as the mean squared Euclidean distance in a 2D plane of target variable pairs: $\vec{v}_i = (\log L_i, \log T_\mathrm{eff,i})$ for the HR diagram and $\vec{v}_i = (\log g_i, \log T_\mathrm{eff,i})$ for the Kiel diagram. This measurement agrees reasonably well with the visual assessment of how closely a predicted track aligns with the true test track. As summary measures of predictive performance on the test data, we take the maximum $L2$ measure, $L2^+ = \max \left\{ L2(M_\mathrm{ini}^\mathrm{test} / \mathrm{M_\odot}) \right\}$ among the 16 initial masses of the test set, for each type of diagram, namely,\  $L2_\mathrm{HR}^+$ and $L2_\mathrm{K}^+$.

\section{Interpolation scheme solutions}\label{sec:ffNN_GPR}

In this section, we describe the methodology behind the development of the two solutions to cost-efficient stellar evolution forecasting over continuous parameter spaces. For the construction of a stellar evolution emulator with supervised machine learning, we treat the selection of the surrogate model class in Sect.~\ref{sec:ML_models}. Then, we discuss loss function choice (Sect.~\ref{sec:loss_func}), and outline our training and hyperparameter optimization methods to obtain the best-fit model (Sect.~\ref{sec:hpo}), which is a feedforward neural network. The hierarchical nearest-neighbor interpolation method is subject of Sect.~\ref{sec:HNNLI}.    

\subsection{Model selection}
\label{sec:ML_models}
There are different surrogate model class candidates available for tackling the regression problem defined in Sect.~\ref{sec:regr}. For selection of statistical learning algorithms, the following three requirements apply in our problem case: 1) applicability to a large data set ($N> 150k$), 2) multiple output\footnote{The multiple output condition (three target variables predicted by a single surrogate model) is motivated by pragmatic considerations: predicting a multitude of stellar variables each with a separate surrogate model requires substantially more effort, if the desired number of output variables of interest is large.}, and 3) fast computational speed in forecast generation, for applicability of the surrogate model at scale. Below, we discuss a number of available options, and justify our choices.

\paragraph{Choice of statistical learning model:} 
GPR has been considered the standard model choice for emulation tasks \citep{Sacks1989}. However, because of memory limitations, the default implementation of global GPR is not applicable to large training data sets. While there are approaches to improve the scalability of GPR, we did not opt for GPR-based emulators for reasons discussed in Appendix \ref{appendix:GPR}. Instead, we tested the performance of a number of regression models that satisfy the aforementioned constraints. After a series of manual tests, we found a satisfactory starting performance with the \textit{k}-nearest neighbors \citep{Fix1989}, random forest \citep{Ho1995}, and feedforward neural network \citep{ivakhnenko1967cybernetics,rumelhart1985learning} regression models classes, all of which are efficient statistical learning algorithms that qualify as scalable predictive models with multiple outputs. Among them, in order to identify which model class is the best choice for the construction of a sufficiently accurate surrogate model, we performed a hyperparameter optimization of each of these three to cross-compare their performance, as assessed by the scores defined in Sect.~\ref{sec:perf}. We performed hyperparameter optimization of \textit{k}-nearest neighbors (KNN) and random forest (RF) regression models by a grid search, with a sampling of numerical hyperparameters over a log scale, and carried out a model selection based on three-fold cross-validation. For the feedforward neural network (ffNN) model, which has a much larger space of options for hyperparameter choices, we determined a preliminary best-fit hyperparameter configuration after training hundreds of models over a high-dimensional, but coarsely sampled hyperparameter grid. We then took it as a starting configuration,  which we further optimized in terms of the hyperparameter selection over a series of manual experiments. The result is that a manually tuned feedforward neural network (ffNN) outperforms KNN and RF models that have been optimized through a grid search, as assessed by the majority of error metrics defined above (see Table~\ref{Tab:table2}). The KNN and RF best-fit models therefore serve us primarily as benchmarks for ffNN performance. 

\paragraph{Deep learning models:} 
ffNN is one out of many available deep learning architectures. We opt for a ffNN architecture because in our regression problem, the input is a vector of fixed dimension. To discriminate, we did not train a recurrent neural network based architecture, which is the model class of choice if the input is a sequence of variable length; nor did we choose a convolutional neural network architecture, which is the model class of choice if the input is a higher dimensional topological data array. A motivation for choosing a ffNN architecture is the established theoretical result that a ffNN with a number of hidden layers $\geq$ 1 is capable of universal function approximation \citep{Hornik1989}. 

\subsection{Choice of loss function}
\label{sec:loss_func}
Choosing a loss function appropriate to the problem is a crucial step because it defines the training goal for supervised machine learning. During the optimization of a ffNN, its trainable parameters are iteratively updated, after each batch, to minimize the loss score. Choosing one error score over another is a trade-off to compromise which type of error is least tolerable against other types of errors. Common choices of scoring rules \citep[for a more detailed reference on scoring rules for point forecast evaluation, see][]{Gneiting2011} for model training as well as for point forecast evaluation are the MAE and MSE. Other choices include the mean squared logarithmic error (MSLE) and the mean absolute percentage error (MAPE). For our problem case, the loss function selection was guided by the following considerations.

MAPE is not the appropriate loss function since, for instance, changes in log-scaled luminosity of massive stars in HR diagram happen on a smaller relative numerical scale than for low-mass stars and prediction errors in that range would therefore hardly be penalized. Furthermore, we chose not to opt for MAPE for additional reasons that are outlined in \cite{Tofallis2015}. When choosing MSLE as loss function, we observed an inefficient learning procedure, with an overly slow decline of MSE, MAE, and our physical performance scores over the validation data. However, we also found neither MAE nor MSE to be optimal choices for our problem. Using MAE allowed the mean averaged error scores to remain low but admitted considerable prediction outliers. Conversely, using MSE reproduced the global shape of the distribution of values of the target variables, but predictions of stellar tracks were often not precise enough locally, and overfitting occurred at epochs much earlier than when minimizing MAE. Instead, we opted for the Huber loss \citep{Huber1964}, which seeks a trade-off between MAE and MSE minimization. It penalizes MSE-like for small prediction errors, and MAE-like for large prediction errors, using the parameter $d$ for the transition threshold \citep[for a recent discussion and generalization, see][]{Taggart2022}:
\begin{equation*}
    L_d ( Y, \hat{Y}) =  \begin{cases}
      \frac{1}{2} ( Y - \hat{Y} )^2& \text{for $ \abs{Y - \hat{Y}} \leq d,$}\\
      d\, \abs{Y - \hat{Y}}-\frac{1}{2} d^2 & \text{otherwise}. 
    \end{cases}
\end{equation*}
During supervised learning, the Huber loss $L_d (Y, \hat{Y})$ issues a penalty for each point prediction error, given the prediction, $\hat{Y,}$ by the surrogate model and the true label, $Y,$ it is compared against. When training deep learning models to predict multiple output, the mean Huber loss is computed as the average across target variables, that is, over the set of labels and over multiple output predictions $\{ \vec{Y}_j, \hat{\vec{Y}}_j\}_{j=1, ..., n_b}$ that are obtained from one randomly sampled data batch of size $n_b$. We find our best results, as assessed by the physically meaningful performance scores outlined in Sect.~\ref{sec:perf}, with $d = 0.75$. 
Once a desired target value of the validation loss score is set, which goes in hand with low enough physical performance scores over the validation data,  what is left is to seek a suitably configured deep learning model that reaches this target value.\footnote{See Appendix \ref{appenxid:AP} for caveats regarding choice of the loss function.} 

\subsection{Hyperparameter optimization}
\label{sec:hpo}
There are two types of hyperparameters that ought to be optimized when constructing ffNN-based emulators: the architecture and learning hyperparameters. The most important architecture hyperparameters are: the number of layers, number of neurons per layer, choice of activation function, and the kernel initialization. The typical important learning hyperparameters are: the learning rate, batch size, choice of optimizer, and the choice of regularization method. There are three different ways to optimize hyperparameters: first, by manual ffNN learning engineering; second, by automated brute-force search methods (for instance, grid or random search); third, by sophisticated search algorithms (e.g., Bayesian optimization or genetic evolutionary search). We opt for manual ffNN learning engineering instead of automated searches, because for deep learning models, the optimal stage when (i.e., at which epoch\footnote{One single epoch is over, once the entire training data set---presented to the network in batch subsets---has been propagated through the network.}) to stop training cannot be faithfully decided  a priori, and it requires a careful consideration of numerical criteria for stopping training if models are optimized in an automated pipeline. Most reliably, it is determined a posteriori by inspection of the fluctuating training and validation data loss curve declines during the runtime. Then, we continue training so long as the degree of overfitting is tolerable. We consider the overfitting to be tolerable so long as the validation loss---even though it may be decaying slower than the training loss at advanced learning stages (i.e.\ at large epoch numbers)---has not reached the flattening plateau stage, nor started to increase. 

\paragraph{Best-fit model:}
For theoretical considerations regarding hyperparameter tuning and the selection criteria we used, the reader is referred to Appendix~\ref{appendix:hyp_tun}. In practice, we found a successful hyperparameter tuning strategy \citep[guided by][]{Goodfellow2017} with the following configurations (see Table~\ref{Tab:best_fit} for a summary). First, a symmetric many-layer (6 hidden layers) architecture with a moderate number of neurons per layer (128), rectified linear unit \citep[ReLU;][]{Hahnloser} activation, \textit{Glorot} uniform \citep[GU;][]{glorot} kernel initialization,  and layer normalization \citep[LN;][]{LN} regularization after each layer. Layer normalization counteracts overfitting while the eight-layer architecture with 128 neurons per hidden layer yields a large enough model capacity to prevent underfitting by over-parametrization. Second, long-term training ($\sim70k$ epochs) at relatively small (512) batch size. Observation of the degree of fluctuation of the loss curves is a means to assess exploration of the high-dimensional trainable parameter space spanned by the biases and by the weighted connections between neurons from neighboring layers in each backpropagation step. The small batch size (as compared to the size of $N'_\mathrm{train}$) adds stochasticity to the learning, and thereby ensures enough exploration, which is aimed to prevent an early flattening of the validation loss curve. Third, a learning rate schedule of slow exponential decay in the \textit{Adam} optimizer \citep{Kingma}: starting with a large enough initial learning rate $\mathrm{lr}_i=10^{-3}$ (to accelerate the gradient descent at beginning stages of learning) and decreasing the learning rate down to a final $\mathrm{lr}_f  \sim5 \cdot 10^{-6}$ toward the end of training (in order to target global rather local minima in the value space of trainable network parameters). The slow gradual decrease is aimed to improve on subtle prediction errors.    

\begin{table}[h!]
  \begin{center}
  \caption{ Summary of loss function choice, architecture and learning hyperparameters adopted for training our best-fit ffNN model, compared to those adopted by \cite{Scutt}.}
    \begin{tabular}{l l l}       
    \hline\hline  
      
      Hyperparameter & Our choice & \cite{Scutt} \\
      \hline
      \# of hidden layers & 6 & 6\\
      \# of neurons per layer & 128 & 64 \\
      Activation function & ReLU & ELU \\
      Kernel initializer & GU &  \\
      Regularization & LN & / \\
      Batch size & 512 & $6 \cdot 10^4$ \\
      Optimizer & Adam & Adam \\
      lr schedule & Exp. decay & Fixed lr \\
      lr range & ($10^{-3}, 5 \cdot 10^{-6}$) & $7 \cdot 10^{-5}$ \\
      Loss function & Huber loss & MSE \\
      \hline
    \end{tabular}
  \label{Tab:best_fit}
  \end{center}  
\end{table}

\subsection{Hierarchical nearest-neighbor interpolation}
\label{sec:HNNLI}
In this section, we present a second method to solve the problem by a HNNI scheme. 
Our construction of the HNNI algorithm was partly motivated by an attempt to customize the operation of the KNN algorithm to our problem setting. In KNN, the nearest neighbors are selected based on a pre-defined distance metric (e.g., Euclidean or Manhattan) over the input parameter space, without treating the regressor dimensions apart from one another. The key principle behind the HNNI method is to select the nearest available grid points (from above and from below) in each parameter space direction to the location in parameter space at which the interpolation prediction is to be made; and then to apply a 1D interpolation prescription subsequently in each parameter space direction according to a hierarchical order of parameters. Our method works similarly to \cite{BrottB} in that it performs a sequence of linear interpolations separately in each parameter space direction according to a hierarchical ordering of stellar variables, but different from it in that it uses a timescale-adapted evolutionary coordinate, instead of fractional age, as the primary interpolation variable. We thereby show that the method is applicable not only to the MS evolution, but to a sequence of evolutionary phases. In this regard, our method is analogous to \cite{Agrawal} in that it uses an adapted evolutionary coordinate to trace the evolution of stars across phases, but we use a prescription for it that allows for automatization of its computation.

We prepare the data set for generating predictions with HNNI under exactly the same conditions as in the supervised machine learning case.
The $N_\mathrm{train}'$ is now used as a catalog data base, upon which the hierarchical nearest neighbor interpolation is performed, instead of serving as the training data for fitting a surrogate model. The HNNI method requires continued access to the pre-computed stellar evolutionary tracks catalog. The HNNI method is applied separately to each of the three surface variables $Y_L$, $Y_T$, and $Y_g$, for obtaining point forecasts at unseen locations in parameter space.

As will be shown in Sect.~\ref{sec:step2_hh_fit}, the HNNI is applicable reliably over the entire initial mass range and over all three evolutionary phases, including the transitions in between them, without the need to map out points of interest for that purpose. The level of predictive accuracy of HNNI is achieved for two main reasons. First, HNNI operates on local parameter space regions immediate to the test location at which a prediction is to be made. Predictions are calculated by an interpolation scheme that treats different dimensions apart from one another. This stands in contrast to the way ffNN, RF, and KNN operate. RF and ffNN take the global properties of the input parameter space into account to find their own rules for making predictions. Comprehension of global patterns can be a great benefit in some problem settings, but irrelevant in others. Similarly to HNNI, KNN also operates on local environments but does not take hierarchy relations among input parameters into account. Second, HNNI uses the normalized timescale-adapted evolutionary coordinate $s$ as primary interpolation variable, without which the interpolation scheme would not produce accurate results. By virtue of using the latter, interpolation-based predictions at transitions between evolutionary phases are mostly accurate because meanwhile values of stellar log-scaled luminosity, effective temperature, or core density variables change drastically. Therefore, the path length increment $\delta s$, which is computed from absolute increments in these variables, increases significantly, resulting in a higher resolution along the age proxy axis of the transition stages between evolutionary phases.\\
Given the initial mass parameter space sampling used in this work, a linear interpolator was sufficient for making accurate forecasts. More generally, for each parameter space dimension, a different  (e.g., a quadratic or cubic polynomial) functional could be applied instead.\\ For clarity, we outline the pseudo-code of HNNI in a 3D $(s, M_\mathrm{ini}, Z_\mathrm{ini})$ single star parameter space in Appendix \ref{appendix:hnni3d}. We believe that the HNNI method, in its basic principle, is applicable to those higher dimensional parameter spaces that allow for a sequential ordering of the parameters in importance of their effect on the shape of resulting stellar evolutionary tracks. 

\section{Results}
\label{sec:results}
In this section, the prediction results, obtained with the deep learning surrogate model and with the HNNI algorithm, are analyzed. We treat the observables fit (step 2) first (Sect.\ \ref{sec:step2_hh_fit}) because it yields the physically meaningful outcome: the prediction of stellar evolution variables and tracks. Therefore, in our two-step interpolation scheme, the observables fit needs to reach a satisfactory level of accuracy first, which can be assessed physically, before approaching the age proxy fit (step 1). Then, the performance baseline for the age proxy fit is set by the condition that the predictive accuracy of the integral two-step interpolation scheme is maintained on the same order of magnitude, as assessed by the scores. We analyze the step 1 fit in Sect.~\ref{sec:step1}. 

\subsection{Observables fit}
\label{sec:step2_hh_fit}

\begin{table}
\caption{Ranking of the predictive models Hierarchical Nearest Neighbor Interpolation (HNNI), feedforward neural network (ffNN), random forest (RF) and \textit{k}-nearest neighbors (KNN) regressors, according to the performance scores outlined in Sect.~\ref{sec:perf} to assess predictive accuracy of stellar observables. The best performance is marked in bold, the worst with a "*" tag. The manually tuned ffNN outperforms the grid search hyperparameter optimized RF and KNN models according to all scores except $\overline{\epsilon}_T$. HNNI outperforms ffNN as assessed by all scores, except $\overline{\epsilon}_L$, $\epsilon^+_g$, and $\epsilon_L^-$.}
\label{Tab:table2}
\centering
\scalebox{0.94}{

\begin{tabular}{c c c c c}        
\hline\hline                 
Score & HNNI & ffNN & RF & KNN\\
\hline                        
\multicolumn{5}{l}{validation data set}\\
$\overline{\epsilon}_L$  & 6.57E-05
  &  \textbf{$-$4.46E-05}
 & 2.612E-04*
 & 2.506E-04
\\
      $\overline{\epsilon}_T$ & \textbf{$-$4.94E-06}
 & 1.82E-04*
 & $-$2.00E-05
 & $-$1.758E-05
\\
      $\overline{\epsilon}_g$ & \textbf{$-$9.22E-05}
 & 4.16E-04
  & $-$4.77E-04*
 & $-$4.55E-04
\\
      $\epsilon_L^{+}$ & \textbf{0.102} & 0.145
 & 0.210
 & 0.217*
 \\
      $\epsilon_T^{+}$ & \textbf{0.014}
 & 0.032 & 0.093
 & 0.095*
 \\
      $\epsilon_g^{+}$ & 0.169 & \textbf{0.165}
 & 1.05
 & 1.08*
\\
      $\epsilon_L^{-}$  & $-$0.115
 & \textbf{$-$0.108}
 & $-$0.700
 & $-$0.721*
 \\
      $\epsilon_T^{-}$  & \textbf{$-$0.011}
 & $-$0.016
 & $-$0.0378*
 & $-$0.0375
\\
      $\epsilon_g^{-}$  & \textbf{$-$0.143} & $-$0.191
 & $-$0.286
 & $-$0.294*
\\
      MSE & \textbf{1.11E-05}
 & 2.01E-05
 & 2.39E-04
 & 5.79E-04*
 \\
      MAE & \textbf{0.00041}
 & 0.00193
 & 0.00479*
 & 0.00270
 \\
 \hline
 \multicolumn{5}{l}{test data set}\\
      $L2_\mathrm{HR}^+$ & \textbf{0.0166}
 & 0.0176 & 0.0319*
 & 0.0237
  \\
      $L2_\mathrm{K}^+$ & \textbf{0.0225} & 0.0283 & 0.0442*
 & 0.0283\\
 \hline

 \end{tabular}
}
 \end{table}

\subsubsection{Deep learning emulation}

\begin{figure*}
  \centering
  \includegraphics[width=0.32\textwidth]{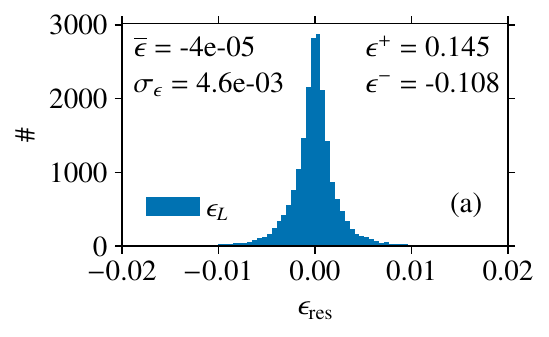}
  \includegraphics[width=0.32\textwidth]{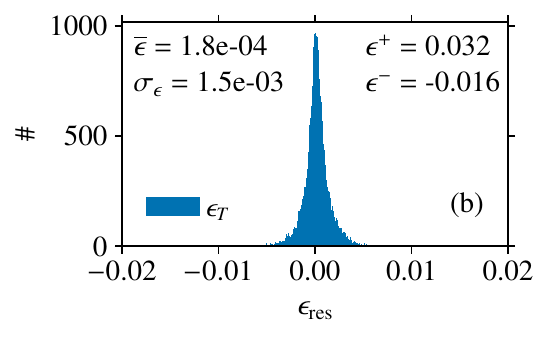}
  \includegraphics[width=0.32\textwidth]{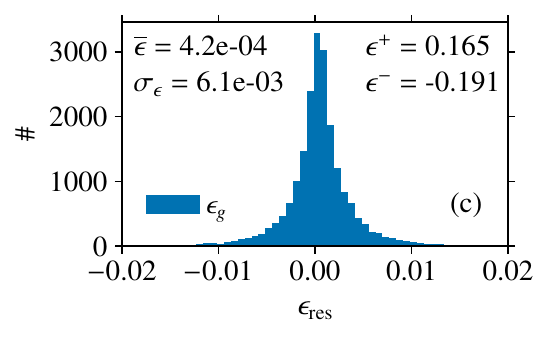}
  \caption{Validation data results for the ffNN-based stellar evolution emulator. The histograms and summary statistics of the residuals $\epsilon_k = Y_k - \hat{Y}_k$, over the validation data $k = 1, \dots, N_\mathrm{val}$ are shown, for $Y_L$ in panel (a), for $Y_T$ in panel (b), and for $Y_g$ in panel (c). 
  We calculate the mean $\overline{\epsilon}$, the standard deviation $\sigma_\epsilon$, the most extremal underprediction $\epsilon^+$, and the most extremal overprediction $\epsilon^-$. Overall, the distribution of residuals is globally symmetric around near 0, with a sharper peak than a Gaussian, reminiscent of a Cauchy distribution.} 
\label{fig:error_stats}
\end{figure*}

\paragraph{Validation data:}
The performance assessment on the validation data is presented in Fig.~\ref{fig:error_stats} by histograms of the residuals and by the summary statistics, defined in Sect.~\ref{sec:perf}, individually for each of the three predicted surface variables. 
If we assume that the prediction errors of $Y_L$, $Y_T$, and $Y_g$ were scored over the same numerical scale, then the following conclusions could be made. The mean residual, in absolute value, is largest for $\log g$ and lowest for $\log L$, while the most extremal overprediction and underprediction are obtained for the $\log g$ target variable. All three mean residuals take on low numerical values on the order of $10^{-4}$ or $10^{-5}$. These error scores are comparable to those found with the best-fit neural network model of \cite{Scutt} ($8 \cdot 10^{-4}$ dex on $\log L$ and $2 \cdot 10^{-4}$ dex on $\log T_\mathrm{eff}$), who address a similar regression problem.
Since $\overline{\epsilon}_\mathrm{res}$ is negative for $\log L$ but positive for  $\log T_\mathrm{eff}$ and $\log g$, the deep learning emulator tends to over-predict the first, but to underpredict the latter two. 
The most extreme prediction outliers are on the order of $10^{-1}$ or $10^{-2}$ in absolute value, namely, up to three orders of magnitude larger than the mean residuals. To better characterize the distribution of errors, we therefore computed an additional score, $\sigma_\epsilon$, which is the standard deviation of the residuals over each target variable. It is a measure of the spread of the prediction errors around the mean residual error, which we find to be on the order of $10^{-3}$ for each of the three target variables.

\paragraph{Comparison to observational uncertainties:}

It is of interest to compare the mean residual errors on the target variables to the typical uncertainties from observations of stars. For stellar bolometric luminosity, the relative error is on the order of $\delta L / L \propto 0.01$ for Gaia observations of solar-like stars \citep{gaia}, which translates into $\delta \log L/\mathrm{L_\odot} = \frac{\delta L}{L} \log e \propto 0.004$. For surface gravity, with $\delta \log g/ \mathrm{[cm \cdot s^{-2}]} \propto 0.1$ \citep[see, e.g.,][]{Ryabchikova}, it is comparatively large. For effective temperature of low-mass stars, the observational error is on the order of $\delta T_\mathrm{eff}/{ \mathrm{K}} \propto 50-100$ depending on stellar class and spectral method \citep{Ryabchikova}. For massive stars, the observational uncertainty on the classical observables typically ranges between $ \delta \log L/\mathrm{L_\odot} = 0.1$, $\delta T_\mathrm{eff} \propto 500 - 2000 \, \mathrm{K}$ and $ \delta \log g / \mathrm{[cm \cdot s^{-2}]} \propto 0.1 - 0.2$ \citep{30doradus, vlt}.

In sum, the mean residual errors on all three target variables are smaller than the typical observational errors on the same log-scaled quantities. In the case of $\epsilon_L$ and $\epsilon_g$, the mean residual errors (note: not only these, but also the expected spreads $\sigma_\epsilon$) are smaller by one to three orders of magnitude depending on statistical score. This means that the prediction errors from the emulator are greater than the observational uncertainties only when the prediction errors belong to the tail of their integral empirical histogram, which comprises cases that are statistically rare. 
For $\epsilon_T$, the histogram of linear-scaled residual errors, $T_\mathrm{eff} - \hat{T}_\mathrm{eff}$, yields a mean residual error of $\simeq 8.3 \, \mathrm{K}$, an expected spread of $\simeq 85 \, \mathrm{K}$, a worst overprediction outlier of $ \simeq 1385 \, \mathrm{K}$ and a worst underprediction outlier of $ \simeq 2885 \, \mathrm{K}$ in absolute values. The expected spread is smaller than the observational uncertainty $\delta T_\mathrm{eff}/\mathrm{K}$ but of a similar order of magnitude. Therefore, inference on effective temperature of low-mass stars using the emulator is (based on the assumption of the aforementioned observational uncertainties) least reliable, out of the three surface variables, in a practical setting.      

\begin{figure*}
  \centering
  \includegraphics[width=0.47\textwidth]{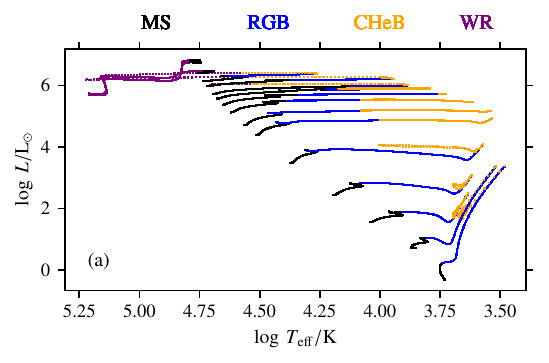}
  \includegraphics[width=0.47\textwidth]{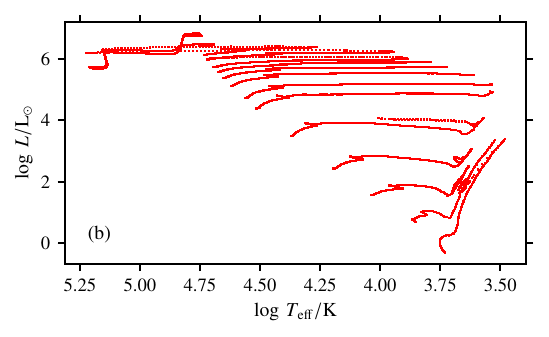}
  \includegraphics[width=0.47\textwidth]{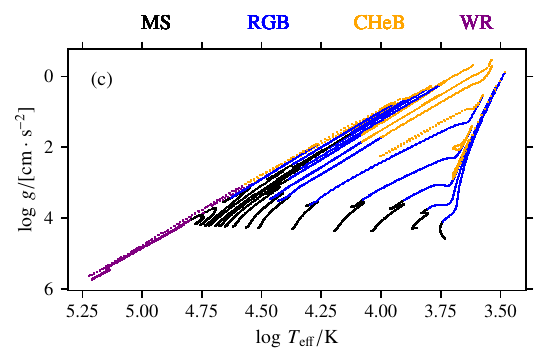}
  \includegraphics[width=0.47\textwidth]{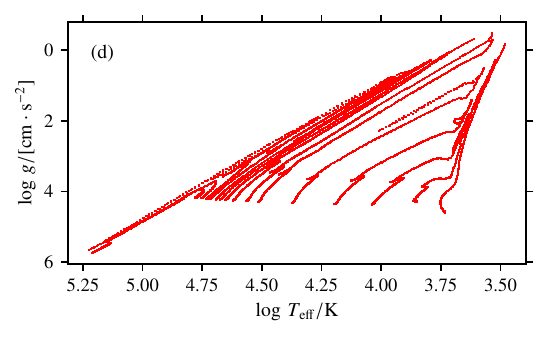}
  \caption{Test data results, comparing the true (left) and the ffNN-predicted (right) stellar evolutionary tracks in HR (top) and Kiel (bottom) diagrams, over the entire set of our test initial masses $\{ M_\mathrm{ini}^\mathrm{test}\}$ unseen by the predictive model during training.}
\label{fig:HR_Kiel_tracks}
\end{figure*}

\begin{figure*}
  \centering
  \includegraphics[width=0.47\textwidth]{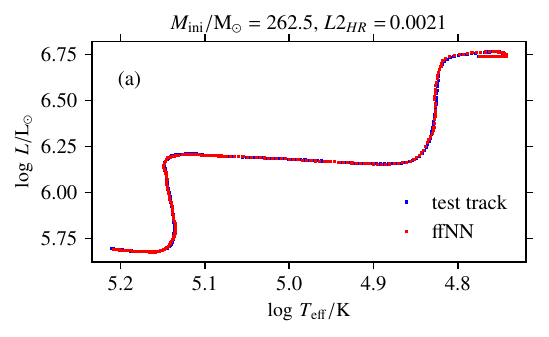}
  \includegraphics[width=0.47\textwidth]{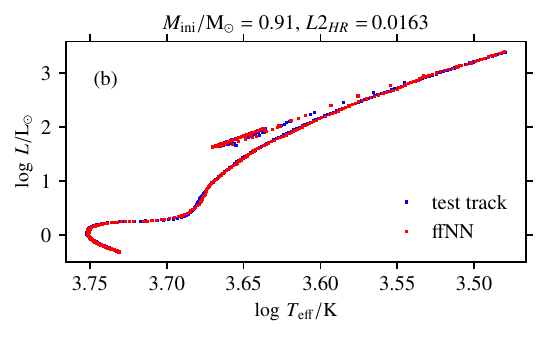}
  \includegraphics[width=0.47\textwidth]{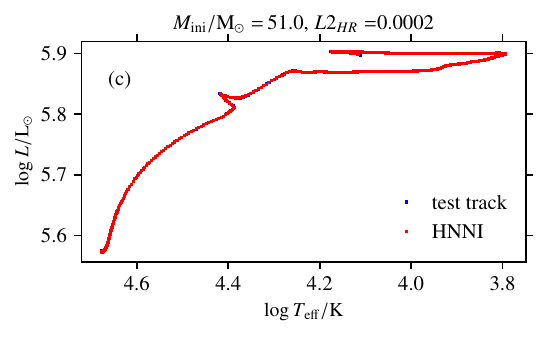}
  \includegraphics[width=0.47\textwidth]{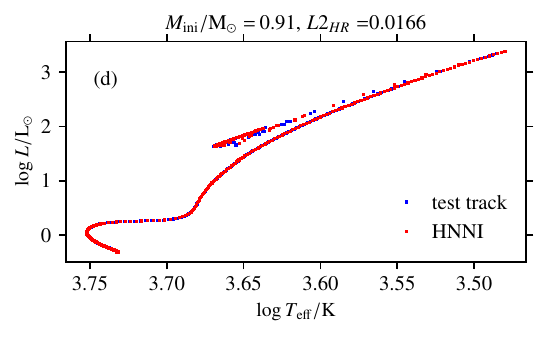}
  \caption{Test data results, showing the best (left) and the worst (right) predictions of stellar evolutionary tracks,  as assessed by the $L2$ measure, in the HR diagram, for unseen test initial masses, by the trained ffNN model (top) and by the HNNI algorithm (bottom). For comparison, the original held-back tracks are underlain.}
\label{fig:worst_best_HR_fit}
\end{figure*}

\paragraph{Test data:}
For model testing on the test data, in order to predict evolutionary tracks in the HR diagram, we compute the values of target variables, $\log \, L_i$, $\log \, g_i$,  and $\log \, T_\mathrm{eff,i}$ at the evolutionary coordinate grid points, $\{ s_i\}_{i=1,\dots, N(M_\mathrm{ini}^\mathrm{test})}$, contained in the held-back series for each test initial mass $M_\mathrm{ini}^\mathrm{test}$. We then plot pairs of predicted target variables against one another to obtain the predicted tracks in the HR and in the Kiel diagram, respectively. These can now be compared with the test data held-back tracks in the diagrams. As shown in Fig.~\ref{fig:HR_Kiel_tracks}, the shape of the stellar tracks is reproduced by the deep learning surrogate models across the entire initial mass range. For a closer inspection of the predictive quality, Fig.~\ref{fig:worst_best_HR_fit} displays the best and worst predictions, respectively, of stellar evolution tracks in the HR diagram at unseen test data initial mass grid points. The biggest deviation between predicted and held-back test stellar track is observed at the low-mass end (worst fit for $M_\mathrm{ini}^\mathrm{test} = 0.91~\mathrm{M_\odot}$). There are two main reasons for this. First, as low-mass stars displace in the HR diagram from ZAMS up to TACHeB, they cover a larger spread in value range of log-scaled luminosity than higher mass stars, due to the stretched-out (in the HR diagram) ascension of the red giant branch. Second, the main contribution to cumulative deviation of predicted to the actual test track for low-mass stars arises during the unstable core helium burning, the sequence of short-lived helium flashes. The helium flashes introduce the most prompt transition in both the $\log L$ and the $\log T_\mathrm{eff}$ variables. Since these are physically uncertain from the modeling perspective, it therefore is not as important to obtain high accuracy prediction of flashes compared to other parts of the stellar evolution track.  We evaluate our state-of-art worst fit as satisfactory, since the more reliable (from the modeling perspective) evolution before and after the flashes is well reproduced by the surrogate model: the evolution up to the tip of the RGB and the stable core helium burning after electron degeneracy in the core is lifted.

\subsubsection{HNNI}
Stellar track predictions in the HR and Kiel diagram are obtained in the same way as described above for the deep learning case. The performance of the HNNI predictive model is assessed using the same data bases, procedures, and metrics as the supervised machine learning models. The outcome is that over the validation and test data, HNNI even outperforms the deep learning method in accuracy of predictions (although not significantly) as is measured by the majority of statistical scores (see Table~\ref{Tab:table2}). Over the test data, the HNNI method yields accurate predictions of stellar evolutionary tracks across the entire initial mass range and over all three evolutionary phases, including the fast-timescale transition regions. For illustration, Fig.~\ref{fig:worst_best_HR_fit} shows the best and worst fit of a stellar track in HR diagram over the test data. 
The HNNI and deep learning models agree on the worst fit for $M_\mathrm{ini}^\mathrm{test} = 0.91~\mathrm{M_\odot}$ for reasons explained above. In the HNNI case, the worst fit is resolved at higher accuracy than in the deep learning case, with a deviation from the test track that is marginal throughout, except during the helium flashes.

Furthermore, the HNNI scheme allows us to predict any stellar evolution variable of interest we tested, by virtue of the same algorithmic prescription for interpolation (see Fig.~\ref{fig:core_density_temp} in the appendix for prediction of stellar-core related variables for unseen test data initial masses). In contrast, by the current setting, the ffNN predicts only those three surface variables which it has been trained upon, as set by the regression problem defined in Sect.~\ref{sec:regr}. In principle, a predictive framework with a large number of time-evolved variables could also be achieved with a ffNN emulator in two different ways. By the first way, the dimension of the output would need to be expanded to match the total number of stellar evolution variables of interest. For example, the values of six stellar variables would be produced as output of the 6 neurons in the outermost layer of the ffNN. However, optimizing such a model by a single globally defined loss score is cumbersome (for a discussion, see Appendix \ref{appenxid:AP}). By the second way, a separate ffNN model with univariate output would need be trained to predict each additional stellar variable of interest. This is the more promising approach out of the two, but requires the construction of a separate hyperparameter-optimized model for each output variable.   

\subsubsection{Method comparison}

\begin{figure}
  \centering
  \includegraphics[width=0.45\textwidth]{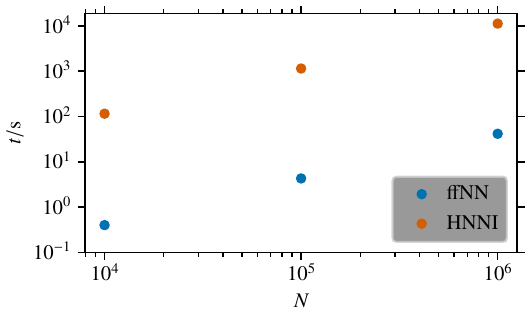}
  \caption{Cost-efficiency of forecast generation. Given our problem size and software implementation of HNNI (see Appendix \ref{appendix:hnni3d}  for an outline of the pseudo-code), the computing time scaling relation $t(N) \propto N$ with the number, $N,$ of multiple output predictions is around 360 times larger for HNNI compared to that of the ffNN.}
\label{fig:efficiency}
\end{figure}

The two methods for stellar evolutionary track forecasting (deep learning emulation versus HNNI) that we develop lie at different trade-off points between cost-efficiency and accuracy of the forecasts. To summarize, the advantages of HNNI are as follows. First, the quality of predictions is reliable, with HNNI even outperforming our best-fit deep learning model. Second, all evolved stellar variables (i.e., not only $\log L_i$, $\log T_\mathrm{eff,i}, \log g_i$, whose prediction has numerically been evaluated for comparison with output of the surrogate model) are covered by the same interpolation prescription. Third, HNNI works as a sustainable out-of-the-box solution method. In contrast to the supervised machine learning approach, there is no need to re-iterate the training and optimization of a predictive model each time another stellar tracks data base is used as the catalog being accessed by the algorithm. \\ The disadvantages of HNNI are as follows. First, continued access to the catalog data base is required, which, depending on the size of parameter space, sampling density and dimension of the problem, is typically of ${\sim} \mathrm{GB}$ size. Second, the computing time to generate predictions is significantly slower compared to the speed of the surrogate model. For the comparison, we have computed scaling relations on a 4-core CPU (see Fig.~\ref{fig:efficiency}): on such a machine, it takes around 40 seconds to generate one million point predictions of all the three surface variables, spread randomly across the evolutionary phases and the initial mass range, with ffNN, while making the same number of predictions takes around 3 hours 13 minutes for HNNI (the computing time scales down linearly with the number of cores that are used to generate the predictions). Third, the extension to higher dimensional parameter space is not straightforward. Depending on the set of stellar parameters, a hierarchical relation may not always be identifiable. Moreover, in a high-dimensional parameter space, the required number of subsequent 1D interpolations becomes large (see the discussion in Appendix \ref{appendix:hnni3d}). Thereby, prediction-generation is slowed down further.      

In contrast, advantages of the supervised machine learning method are as follows. First, casting predictions is fast, namely, two orders of magnitude faster (in seconds) to generate than with HNNI. Second, trained surrogate models are handy: a predictive ffNN model is of file size ${\sim}3$ MB. Third, the supervised machine learning approach is very general: the extension to higher dimensions (in contrast to HNNI) does not require any hierarchical ordering of regressor variables, nor does casting predictions face any significant increase in computing time with increasing dimension. \\The disadvantages of the method are as follows. First, the optimization of deep learning models is a more entailed task than a hard-coding adjustment of HNNI. Second, minimizing a single global loss score during model training does not guarantee locally accurate fit results consistently over the entire parameter space (see Appendix \ref{appenxid:AP} for a discussion thereof and proposed solutions). Third, the scaling of ffNN output with the number of target variables either comes under sacrifice of predictive accuracy (in the multiple output case) or implies considerably more development effort (in the single output case). 

\subsection{Age proxy fit}
\label{sec:step1}
\begin{figure*}
  \centering
  \includegraphics[width=0.32\textwidth]{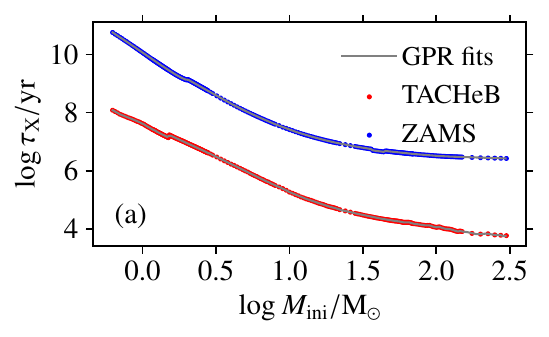}
  \includegraphics[width=0.32\textwidth]{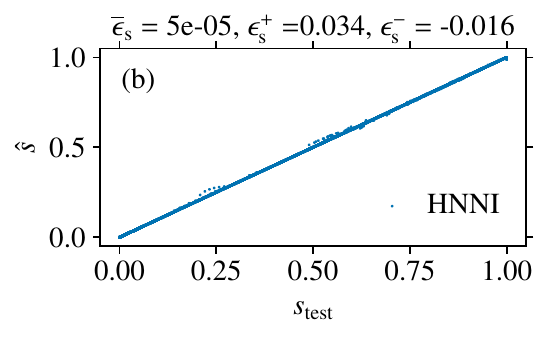}
  \includegraphics[width=0.32\textwidth]{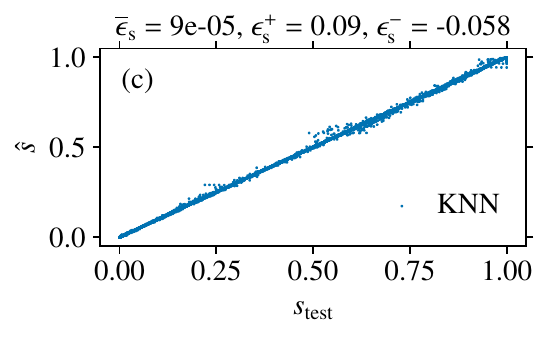}
  \caption{GPR fits of the $\log \tau_\mathrm{ZAMS}(\log M_\mathrm{ini}/\mathrm{M_\odot})$ and $\log \tau_\mathrm{TACHeB}(\log M_\mathrm{ini}/\mathrm{M_\odot})$ relations (a), and the scatter plots of the age proxy predictions $\hat{s}(\log \tau, \log M_\mathrm{ini})$ against the validation data, $s_\mathrm{test}$, obtained with the HNNI (b) and KNN (c) methods, for performance comparison.}
\label{fig:age_n_scatterplots}
\end{figure*}

 The series of age proxy values from $s=0$ (ZAMS) to $s=1$ (TACHeB) are not known for initial masses over which no stellar evolution tracks have been pre-computed, since $s$ is calculated from the $\log L$, $\log T_\mathrm{eff}$, and $\log \rho_c$ time series which are then not available at those initial mass grid points. Both our methods for predicting stellar evolution tracks rely on the timescale-adapted evolutionary coordinate, $s$, which we use to re-parametrize the evolution of stars.
Many astrophysical applications, however, require an indication of the stellar ages; for instance, drawing model isochrones into observed color-magnitude diagrams. We therefore construct another duet of interpolation methods (with HNNI and with supervised machine learning) that map the age $\tau$ onto the value of a star's timescale-adapted evolutionary coordinate $s(\log \tau, \log M_\mathrm{ini}) \in (0, 1)$ over a continuous initial mass range, in order to accomplish the two-step interpolation scheme as defined in Sect.~\ref{sec:regr}. Time counting in the MIST data set starts with the pre-MS phase. Therefore, the values of ages at ZAMS, $\tau_\mathrm{ZAMS}(M_\mathrm{ini})$,  quantify its duration. Instead of $\tau$, we use a scaled age variable $\tilde{\tau} \in (0, 1)$  
for the age proxy fit with both the HNNI and the supervised ML methods: 
    \begin{equation*}
        \tilde{\tau}_i(M_\mathrm{ini}) = \frac{\log \tau_i(M_\mathrm{ini}) - \log \tau_\mathrm{ZAMS}(M_\mathrm{ini}) }{\log \tau_\mathrm{TACHeB}(M_\mathrm{ini})-\log \tau_\mathrm{ZAMS}(M_\mathrm{ini})}.
        \label{eq:scaled_age}
    \end{equation*}    
To obtain back the actual non-normalized age values (in units of years), the supply of the ZAMS  $\log \tau_\mathrm{ZAMS}(M_\mathrm{ini})$ and the TACHeB $\log \tau_\mathrm{TACHeB}(M_\mathrm{ini})$ functions is needed. The $\tau_{ZAMS}(M_\mathrm{ini})$ values are available from the MIST data set at the discretely sampled initial mass grid points. In order to be able to predict ZAMS and TACHeB ages of stars over a continuous $M_\mathrm{ini}$ range, we fit a Gaussian process regression model 
to the discretely sampled catalog ZAMS and TACHeB grid points (see Fig.~\ref{fig:age_n_scatterplots} a), respectively.

\begin{figure*}
  \centering
  \includegraphics[width=0.99\textwidth]{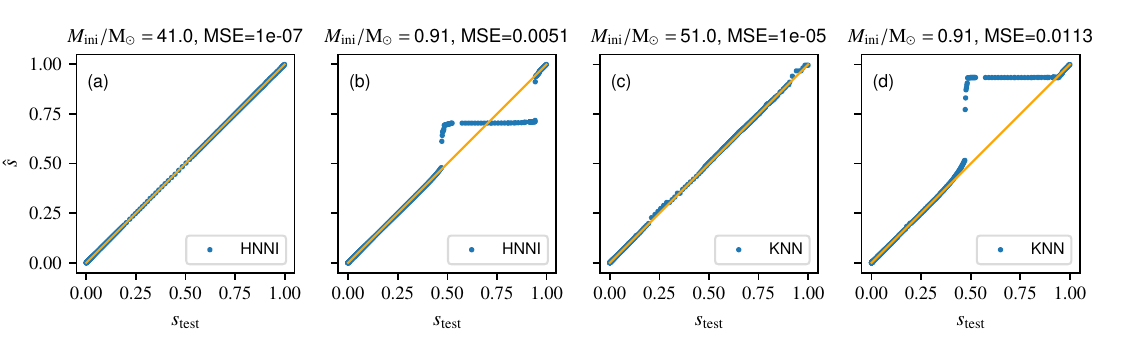}
  \caption{Best [(a) and (c)] and worst [(b) and (d)] fits of age proxy tracks for unseen test initial masses, with the HNNI and KNN methods, respectively.}
\label{fig:step1bestworst}
\end{figure*}

\subsubsection{HNNI}
The HNNI routine for the age proxy fit operates in the same way as outlined in Sect.~\ref{sec:HNNLI}, with the sole difference that the primary regressor variable now is $\tilde{\tau}$ (instead of the age proxy used in step 2), while $s$ is itself the target variable of the fit. As Fig.~\ref{fig:age_n_scatterplots} (b) shows, HNNI predicts the values of the age proxy reliably throughout evolution of stars from $s=0$ up to $s=1$ over the validation data set. The mean residual error $\overline{\epsilon}_\mathrm{res}$ is of order $10^{-5}$. The only clustered scatter regions off the diagonal are around $s \simeq 0.25$ and $s \simeq 0.6$, but the scatter offsets are low in amplitude. The most extremal over- and underprediction outliers are of order $10^{-2}$ in absolute value.\\
The performance evaluation based on the test data assesses the predictive accuracy of mapping the stellar age onto the timescale-adapted evolutionary coordinate over the course of the entire evolution from ZAMS up to TACHeB for unseen initial masses.\\ 
The outcome is that HNNI predicts the value of the age proxy reliably throughout the evolution from ZAMS to TACHeB except at fast-timescale transitions, which make up little stellar lifetime but manifest themselves as sharp increases in age proxy values as function of age.  Fig.~\ref{fig:step1bestworst} (a and b) show the best and the worst fit, respectively, assessed by the MSE metric, among all the test initial masses. The scatter plot of the best fit (for $M_\mathrm{ini}^\mathrm{test} = 41~\mathrm{M_\odot}$) has no considerable spread, since $\hat{s}$ aligns with $s_\mathrm{test}$ over course of the full evolution. In the scatter plot of the worst fit (for $M_\mathrm{ini}^\mathrm{test}  = 0.91~\mathrm{M_\odot}$), the local deviations of $\hat{s}$ from $s_\mathrm{test}$ are apparent: the age proxy is first reproduced accurately up to $s \simeq 0.47$. Then a gap in the output range forms, such that the next predicted age proxy value is at $s \simeq 0.6$ and continues to be overpredicted up to $s \sim 0.7,$ from where  it transitions onward to an underprediction phase. A second domain gap forms and the predicted age proxy aligns back with its actual test data value at $s \sim 0.9$ up to the end. The physical implications on the prediction of HR and Kiel diagrams in effect of the two-step interpolation scheme are discussed in Sect.~\ref{two-s}. Age proxy prediction errors imply that the evolutionary state of the star is either under- or overpredicted, since a wrong evolutionary coordinate value has been assigned to a given stellar age of reference. However, as has been shown in Sect.~\ref{sec:step2_hh_fit}, sampling target variables at homogeneously distributed $\delta s$ increments  (e.g.,\ an equidistant spacing)  in the step 2 scheme ensures that no significant changes in target variability will have been jumped over. In other words, artifact gap formation along curves in HR and Kiel diagrams is avoided by the fitted age proxy parametrization of stellar evolutionary tracks (step 2 fit), independent of the age proxy forecasts (step 1 fit). 

\subsubsection{KNN}
Analogously to the procedure for the observables fit, we constructed another solution to the age proxy fit with supervised machine learning. The reason is that we would like to obtain a more cost-efficient interpolation model than HNNI, which nevertheless is sufficiently accurate for astrophysical application purposes. The age proxy fit is a univariate regression problem distinct from the observables fit and for which the procedure of surrogate model class selection and hyperparameter optimization needs be re-iterated. For comparing and selecting ML surrogate models, we used performance scores that are defined analogously to the performance scores for the step 2 fit (Sect.~\ref{sec:perf}), but applied to the univariate output of age proxy prediction. After a series of tests of a number of model classes, including ffNN, we obtained the best performance with the KNN algorithm. After a preliminary grid search for hyperparameter optimization of KNN, we manually fine-tuned hyperparameters for best-fit results. We obtained these with two neighbors to query, a Minkowski metric, a $p=2$ power parameter for distance calculation, the \textit{BallTree} algorithm, distance-based weighting, and a leaf size of 300. The predictive quality is lower, but error scores are on the same order of magnitude compared to the HNNI case (see Fig.~\ref{fig:age_n_scatterplots}c for the summary statistics of the age proxy prediction errors over the validation data). Therefore, we evaluated the solution with the KNN algorithm as sufficiently accurate. For a performance assessment over the test data set, Fig.~\ref{fig:step1bestworst} (c and d) show the best and worst fits of the age proxy, respectively. HNNI and KNN agree on the worst fit to be at the low-mass end, for $M_\mathrm{ini}^\mathrm{test} = 0.91~M_\odot$. Here, the KNN worst-fit has a characteristic similar to the HNNI case: the age proxy is first reproduced accurately up to $s \simeq 0.52$. Then a gap in the output range forms, such that the subsequent predicted age proxy value is at $s \sim 0.76$. The gap is larger than in the HNNI case. Hereafter, the age proxy is overpredicted, and aligns with the $s_\mathrm{test}$ values from $s \sim 0.93$ onward up to the end.

\subsection{Predicting stellar evolution}
Consecutively putting  together two predictive models for the age proxy and for the stellar observables, respectively, allows for the prediction of stellar evolution tracks in HR and Kiel diagrams as function of stellar age, and of isochrones showing stars of same age. In this section, we use the integral two-step interpolation scheme to predict complete stellar evolution tracks in HR and Kiel diagrams over the set of test initial masses (see Sect.~\ref{two-s}) and to predict stellar observables at fixed values of stellar age over a densely sampled initial mass range (see Sect.~\ref{sec:iso}).  

\subsubsection{Evolutionary tracks}
\label{two-s}
\begin{figure*}
  \centering
  \includegraphics[width=0.49\textwidth]{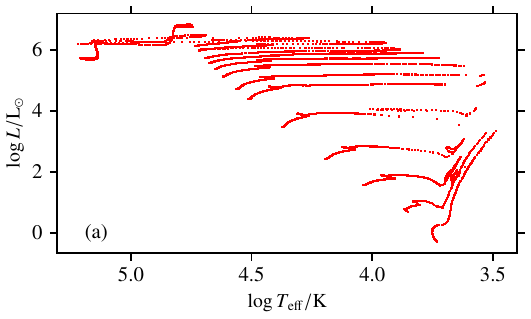}
  \includegraphics[width=0.49\textwidth]{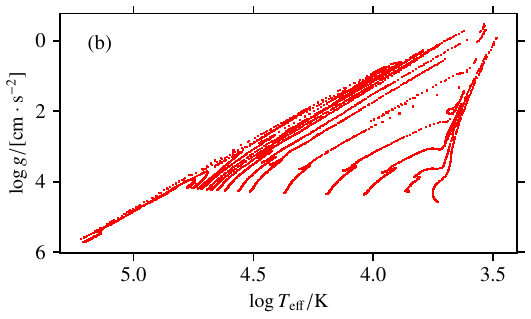}
  \includegraphics[width=0.49\textwidth]{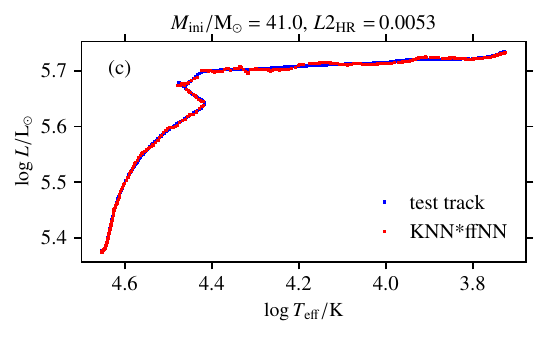}
  \includegraphics[width=0.49\textwidth]{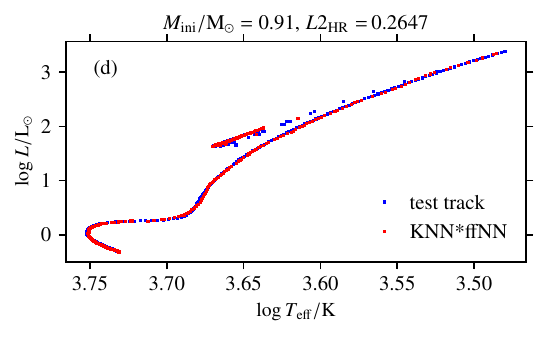}
  \caption{Outcome of the two-step interpolation scheme with supervised ML. Stellar tracks in the HR (top left) and Kiel diagram (top right) for unseen test initial masses are predicted as function of age $\tau$. 
  For comparison with the true test HR and Kiel tracks, see Fig.~\ref{fig:HR_Kiel_tracks}. For better visibility, the best- (bottom left) and worst-fit (bottom right) of tracks in the HR diagram, as assessed by the $L2$ measurement, are displayed separately.}
\label{fig:err_prop}
\end{figure*}

For the input of age ($\log \tau$) and initial mass ($\log M_\mathrm{ini}$) of the star, the value of the age proxy ($\hat{s}$) is predicted first by the step 1 method. Then, $\hat{s}$ is used as input variable for the step 2 method, together with again the initial mass ($\log M_\mathrm{ini}$). Here, we present the two-step pipeline interpolation results that are obtained with the supervised machine learning models (KNN and ffNN), which is a less accurate method compared to HNNI in both fitting tasks. We find that the predictive quality of stellar surface variables reaches the desired accuracy level (see the predictions of evolutionary tracks in HR and Kiel diagrams for unseen test initial masses in Fig.~\ref{fig:err_prop}). 
The net effect the step 1 fit errors have on the shape of predicted tracks in HR and Kiel diagrams in the two-step interpolation scheme is that step 2 based point forecasts of surface variables are incorrectly shifted along the track. If the step 1 fit by KNN accomplished a perfect $\log \tau$-to-$s$ mapping, then the tracks predicted by ffNN (step 2 fit) would retain the same shape as shown in Fig.~\ref{fig:HR_Kiel_tracks} (b and d). However, as the step 1 fit introduces over- and underprediction errors of the age proxy values, these lead to locally increased or decreased sampling densities of the age proxy axis as compared to the original unperturbed case. The under-densities along the age proxy axis result in domain gaps, over which no corresponding step 2 output (values of surface variables) is produced. These gaps form predominantly at fast-timescale transitions between evolutionary phases. As can be seen in Fig.~\ref{fig:err_prop}, this applies in particular to the rear part of passage through the Hertzsprung gap and toward the late stages of CHeB (for high-mass stars), to the nearing of the tip of the RGB and during the helium flashes (for low-mass stars). Depending on accuracy or sampling needs of specific astrophysical applications, post-processing methods could be applied to fill the prediction gaps. The post-processing method would need to identify the domain gaps in the age proxy value range, sample the age proxy within the gap regions to obtain the prediction of observables (by the model that accomplishes the step 2 fit, which is ffNN or HNNI) and then use local interpolation-based methods to infer the stellar ages over the gap regions along the age proxy axis. 

\subsubsection{Isochrones}
\label{sec:iso}

\begin{figure}
  \centering
  \includegraphics[width=0.5\textwidth]{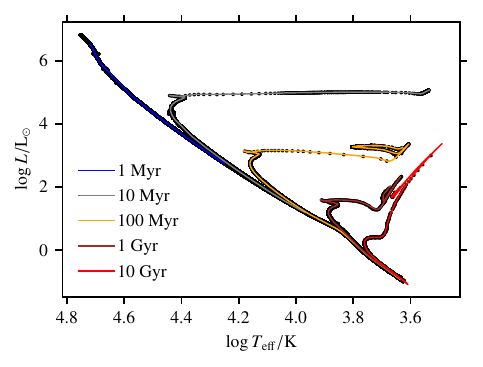}
  \caption{
  Comparison of MIST isochrones with emulator-based (GPR, KNN and ffNN) predictions of stellar observables at fixed ages. The initial mass range $M_\mathrm{ini } \in (0.65, 300) \, \mathrm{M_\odot}$ is sampled over a log scale with a step size $\delta \log M_\mathrm{ini}/ \mathrm{M_\odot} = 5 \cdot 10^{-5}$ to obtain the parameter space points at which discrete predictions are made. For the set of values of stellar isochrones, we chose a log sampling of the time axis to cover the full range of stellar ages $\tau$. The theoretical MIST isochrones are color-marked, while the emulator-based point predictions are scatter-plotted in black.}
\label{fig:iso_plot2}
\end{figure}

Finally, we further demonstrate consistency of our predictive models with the original MIST stellar evolution catalog by comparison of MIST isochrones with emulator-based predictions of observables at fixed stellar ages. MIST isochrones, interpolated over
the parameter space using the EEP-based method discussed in Sect.~\ref{sec:introduction},  are imported from the Web interface\footnote{\url{https://waps.cfa.harvard.edu/MIST/interp_isos.html}} provided by the makers of the catalog. For the emulator isochrones, we use the following multistep predictive pipeline to obtain stellar observables at fixed age:\\ 
First, for each $M_\mathrm{ini}$ value of interest, the two fitted GPR models are used to predict $\log \tau_\mathrm{ZAMS}(M_\mathrm{ini})$ and  $\log \tau_\mathrm{TACHeB}(M_\mathrm{ini})$. Second, these are then used to calculate analytically the scaled age $\tilde{\tau}$ for each pair \{$M_\mathrm{ini}$, $\tau$\} of interest. Third, together with $\log M_\mathrm{ini}$, $\tilde{\tau}$ serves as input of the trained KNN model, which predicts the corresponding value of the timescale-adapted evolutionary coordinate $s$. Fourth, $s$ and  $\log M_\mathrm{ini}$ serve as inputs for the trained ffNN model to predict the observables $\log L$, $\log T_\mathrm{eff}$, and $\log g$, which are then plotted against one another.\\ 
Fig.~\ref{fig:iso_plot2} shows the outcome of the multistep predictive pipeline to obtain isochrones.\footnote{The value range of the imported MIST isochrones is adapted to match our problem setting: only the evolutionary sequences ZAMS--TACHeB are shown, the pre-MS and the post-CHeB evolution of stars are cut off. Note that in addition, there is an intrinsic cut-off in the MIST isochrones at the high-mass end, which --in contrast to the MIST training data set we used-- do not include the WR stars. Therefore, the emulator-based 1 Myr isochrone extends further to the blue part of the HR diagram than the MIST isochrones.} 
For each value of the stellar age, the distribution of point predictions in the HR diagrams mimics a simulated observation of stars under assumption of a log-uniform initial mass probability distribution. Therefore, most emulator-based point predictions of observables populate those regions of stellar evolution tracks over which stars evolve on the nuclear timescales. While there is some scatter of emulator-based predictions  about the theoretical isochrones, most prominent at the blue end of the HR diagram and along the blue loop of the 100 Myr isochrone, we consider the overall statistical match as satisfactory. 

\subsubsection{Comparison with observations}

As has been stressed in Sect.~\ref{sec:introduction}, in this work, we use the MIST catalog for a proof-of-concept study to demonstrate our method of constructing accurate predictive models of stellar evolution over a width of parameter space necessary for a scalable general-purpose astrophysical applicability. We therefore proceed with the background assumption that the MIST data set is the ground truth of stellar evolution modeling. In Sect.~\ref{sec:step2_hh_fit} we show that the emulator-based prediction errors on all the three observables $\log L$, $\log T_\mathrm{eff}$, and $\log g$ are significantly lower than typical observational uncertainties on the same variables. However, our predictive models can explain observations only as good as the original MIST models do. The important question of how well MIST stellar models agree with the observations and which sources of systematic uncertainty have been identified, is addressed elsewhere, namely, in the original paper on the catalog \citep{Choi2016} and in follow-up studies.  In this section, we provide a brief summary of their main conclusions concerning the ZAMS--TACHeB evolution at solar metallicity over the initial mass range $(0.65, 300) \, \mathrm{M_\odot}$. It is aimed at informing interested readers about which scalable astrophysical applications of our predictive models are reasonable, and which are not, as consequence of systematic prediction errors that result from the adopted MIST input physics. Our trained machine learning models can be used for astrophysical applications in future work and are available to corresponding authors upon reasonable request.\\
MS evolution, MS turn-off morphologies, and red clump luminosities of low-mass stars are in good agreement with MIST predictions, except for those in the mass range $M_\mathrm{ini} < 0.7 \, \mathrm{M_\odot}$. The MS evolution of high-mass stars is reproduced well by MIST models close to ZAMS, but not the MS width at the highest masses within the tested range $M_\mathrm{ini}/ \mathrm{M_\odot} \in (10, 80)$. The slope of model red supergiants is too shallow compared to observations; however, no observed red supergiants lie in the forbidden zone cooler than the limit at the Hayashi line. Comparing the observed to predicted ratio of WC- to WN-type\footnote{ WC stars are WR subtype stars that reveal helium-burning products in the outer layers, while WR stars of subtype WN reveal hydrogen-burning products.} stars, of WR to O-type, and of blue- to red supergiant stars allows us to test mass loss, semiconvection, and convective overshoot models. At $\mathrm{Z_\odot}$, model ratios and observed ratios broadly agree on the order of magnitude, but the deviation is substantial in particular for the ratio of WC- to WN-type stars. For a more detailed analysis, the reader is referred to the original paper \citep{Choi2016} and references therein.  

\section{Conclusions and outlook}\label{sec:conclusions}
\label{sec:concl}

We develop two method classes for interpolation of stellar evolution tracks over an initial mass range from red dwarfs to very massive stars, evolved from the zero age main sequence (ZAMS) up to terminal age of core helium burning (TACHeB). The two interpolation methods are: 1) a surrogate model of stellar evolution constructed with supervised machine learning and 2) a catalog-based hard-coded hierarchical nearest-neighbors interpolation (HNNI) algorithm. Both of these invoke a two-step interpolation procedure that makes use of a timescale-adapted evolutionary coordinate $s$ (age proxy) that we introduce to re-parametrize the evolution of stars. This re-parametrization reduces the timescale variability of evolutionary variables, thereby allowing for more accurate predictions across timescale-separated evolutionary phases.\\ 
For the predictive two-step pipeline constructed with supervised machine learning, we optimized a \textit{k}-nearest neighbors model to predict the age proxy for the input of (scaled) stellar age, $\tilde{\tau,}$ and initial mass, $M_\mathrm{ini}$. The predicted age proxy value, together with initial mass, is then used as input by a hyperparameter-tuned feedforward neural network model to produce the multiple output prediction of the log-scaled surface variables luminosity, $\log L$, effective temperature, $\log T_\mathrm{eff}$, and surface gravity, $\log g$. These predictions allow tracing the evolution of stars in the HR and Kiel diagrams over the dominant duration of their lifetimes.\\ 
For the predictive two-step pipeline constructed with HNNI, we use the same syntax in the algorithmic prescription for both the age proxy prediction and for the prediction of observables. It operates by selecting the two nearest neighbors, from above and from below, in each parameter space direction, and then performing a sequence of linear interpolations, according to a hierarchical ordering of parameters.\\
Depending on the astrophysical application, one method is preferable over the other. The supervised machine learning approach is more cost-efficient (by two orders of magnitude in seconds) but more difficult to develop. The hard-coded HNNI is more accurate by one order of magnitude on the MAE and by two orders of magnitude on $\overline{\epsilon}_T$, while all other error scores are on the same scale), but less handy, since continued access to the stellar evolution catalog is required.\\ 
With a wide initial mass range and with a sequence of evolutionary phases from the ZAMS up to TACHeB,  astrophysical application of our models is of interest, first, in context of rapid single star population synthesis. The second promising application prospect is the incorporation of stellar evolution emulators as stellar microphysics sub-grid models in large-scale stellar $N$-body dynamics or galactic evolution simulations. The third prospect for application  is the usage  of our interpolation methods to infer fundamental stellar parameters (given multiple observables of a single star) or the initial mass function (given the observation of a stellar population). The latter astrophysical application prospect could follow the \textsc{Dalek} \citep{Kerzendorf2021} working example in context of spectral modeling of Type Ia supernovae. \textsc{Dalek} is a deep learning based emulator of the output of the \textsc{Tardis} \citep{Tardis2014} radiative transfer code. 
A variant thereof has been used in a Bayesian framework, where it represents the output of \textsc{Tardis}, to infer supernova progenitor parameters from the observation of its spectrum \citep{OBrien2021}. The variant has been trained from scratch on a training data set which has been generated over a reasonably constrained parameter range  given likely properties of the progenitor system.  Reliable inference on the parameters of the progenitor system without the emulator, with the traditional grid-based methods instead, is impossible; it would take thousands of years of clock time to evaluate the high-dimensional parameter space by the classical Bayesian inference approach of running \textsc{Tardis} models at all those parameter space grid points as selected by iterative optimization that typically requires millions of evaluations.\\
Sampling a stellar evolution track as function of the age proxy instead of the stellar age, for instance at equidistant $\delta s$ increments, facilitates the adequate resolution of all significant changes in the stellar output variables.
This applies not only to the ZAMS--TACHeB sequence, but also to the pre-MS and post-CHeB evolution (up to white dwarf cooling for low-mass stars).\\ 
For the generalization of our methods to a higher dimensional space of fundamental parameters, additional considerations need to be taken into account. Sampling of a high-dimensional parameter space to generate the grid data needs to be efficient: sparse enough to keep the computational expenses low, but dense enough to maintain the predictive accuracy satisfactory locally across all directions in parameter space. The MIST single star grid space sampling density distribution, which we used to construct our models, has been decided upon by the makers of the catalog, based on physical insight from domain expertise. We have expanded the data set in parameter ranges of interest based on inspection of local fit results  obtained with the surrogate model, to locally improve the predictive performance where needed, by supplying more training data in those regions. An alternative approach to determining the optimal parameter space sampling goes by using active learning \citep[AL;][]{Settles2009ActiveLL}. By pre-defined heuristics, decision-making with AL is automated and, therefore, better adapted to high dimensional parameter spaces for finding an optimized distribution of grid points.  In the context of stellar astrophysics, \cite{Rocha2022} applied AL in a case study involving the mapping of initial binary star parameters to the final orbital period and show that it can be used to reduce the training data grid size. \\
For stellar parameter spaces greater than those tested here, we recommend using HNNI as the predictive interpolation model as far as it is applicable given computational cost constraints. The HNNI method generalizes to higher dimensions: for clarity, we have provided the recipe for a 3D ($s, M_\mathrm{ini}, Z_\mathrm{ini}$) formulation of HNNI in order to show the systematic of its dimensional extension. 
In the case that either the HNNI method we developed will break down or be computationally too inefficient in the high-dimensional parameter space (given the impractically large cumulative number of 1D interpolations to make), we recommend using supervised machine learning, in particular deep learning, to train univariate surrogate models of stellar evolution on segments of the initial mass parameter space. For training deep learning models, we have provided basic guidance on selection of feedforward neural network architecture and learning hyperparameters and on the choice of the loss function. Finally, we have found a successful training strategy that, in its basic design, could (since it has been adjusted to data base specifics of a stellar evolution catalog) continue to produce satisfactory fit results when trained on data in a higher dimensional parameter space.
\begin{acknowledgements}
We thank Tilmann Gneiting, Giuliano Iorio, Saskia Hekker, Ralf Klessen, Frauke Gr\"ater, Sebastian Lerch and Evgeni Ulanov for helpful advice and discussions, and the anonymous referee for useful suggestions. The authors acknowledge support by the Klaus Tschira Foundation. This work has received funding from the European Research Council (ERC) under the European Union’s Horizon 2020 research and innovation programme (Grant agreement No.\ 945806). This work is supported by the Deutsche Forschungsgemeinschaft (DFG, German Research Foundation) under Germany’s Excellence Strategy EXC 2181/1-390900948 (the Heidelberg STRUCTURES Excellence Cluster).
\end{acknowledgements}
  

\bibliographystyle{aa}
\bibliography{biblio}

\begin{appendix}

\section{Stellar evolution re-parametrization and HNNI}
\begin{figure*}
      \centering
      \includegraphics[width=0.32\textwidth]{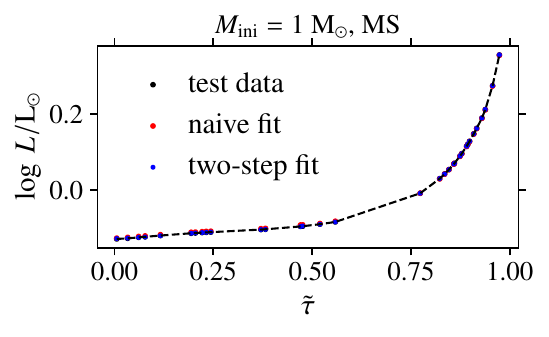}
      \includegraphics[width=0.32\textwidth]{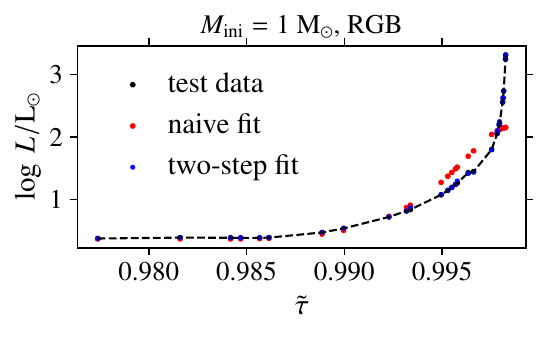}
      \includegraphics[width=0.32\textwidth]{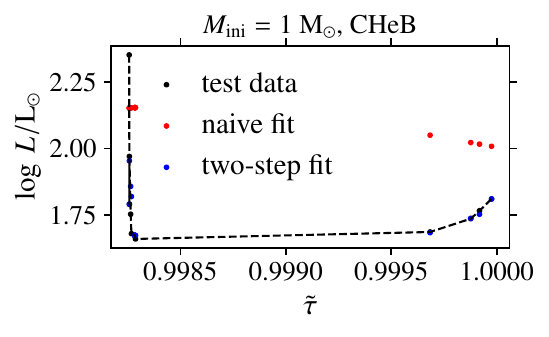}
      \caption{Comparison of naive versus a two-step fit of stellar evolution time series, upon the 1D test case of predicting log-scaled luminosity of a Sun-like star over the MS, RGB, and CHeB phases.} 
    \label{fig:naive-twostep}
\end{figure*}

\begin{figure*}
  \centering
  \includegraphics[width=0.49\textwidth]{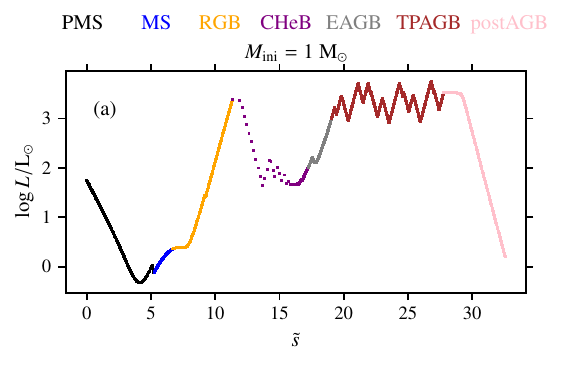}
  \includegraphics[width=0.49\textwidth]{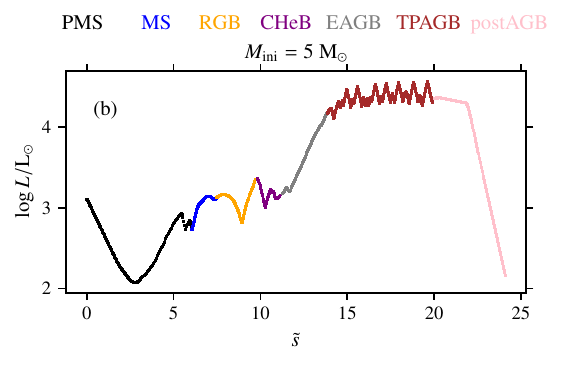}
\includegraphics[width=0.49\textwidth]{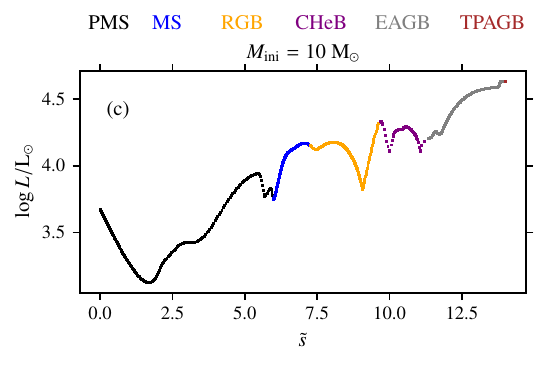}
  \includegraphics[width=0.49\textwidth]{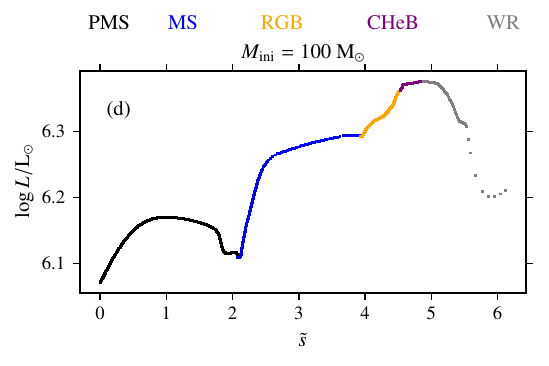}
  \caption{Parametrization of stellar evolution as function of the timescale-adapted evolutionary coordinate over phases beyond the ZAMS--TACHeB sequence. Stars across the initial mass range are evolved from the pre-MS up to post-CHeB evolution (post-asymptotic giant branch evolution for low-mass stars, and onset of core carbon burning for massive stars) as function of (unnormalized) timescale-adapted evolutionary coordinate $\tilde{s}$. The color marking denotes the evolutionary phases pre-main sequence (PMS), main sequence (MS), red giant branch (RGB), core helium burning (CHeB), early asymptotic giant branch (EAGB), thermally pulsating asymptotic giant branch (TPAGB), post asymptotic giant branch (postAGB), and Wolf-Rayet (WR) phase.}
\label{fig:age-proxy2}
\end{figure*}

\begin{figure*}
  \centering
  \includegraphics[width=0.49\textwidth]{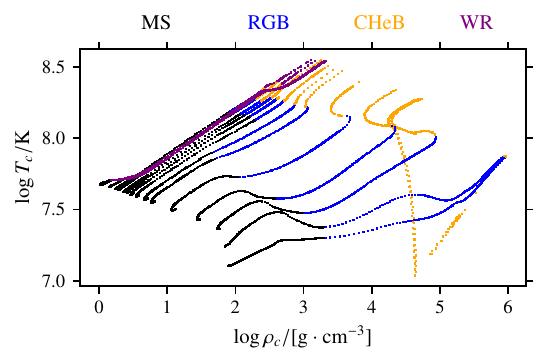}
  \includegraphics[width=0.49\textwidth]{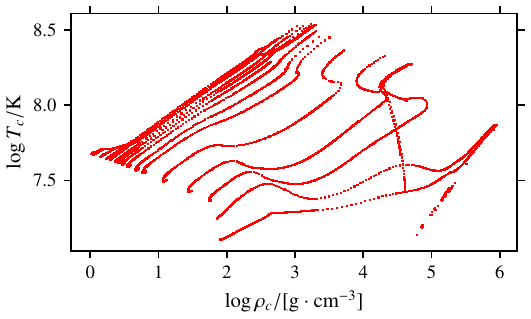}
  \caption{Prediction of stellar evolutionary tracks in the core density-temperature diagram over the test data, with the HNNI method (right) and the true held-back tracks (left) for comparison.}
\label{fig:core_density_temp}
\end{figure*}

Figures \ref{fig:naive-twostep}--\ref{fig:core_density_temp} provide additional materials that support the use of a timescale-adapted evolutionary coordinate and of the HNNI method for interpolation of stellar evolution tracks. Fig.~\ref{fig:naive-twostep} demonstrates the general suitability of our age proxy prescription for resolution of variability in stellar tracks over a wide span of sequential evolutionary phases and across the initial mass range. \\ Fig.~\ref{fig:age-proxy2} compares the predictive quality of the two-step fitting approach with the naive direct fit, when applied to the test case of modeling the log-scaled luminosity time series of a Sun-like star from ZAMS up to TACHeB. We make an 85:15\% train-test split of the data and use the scaled age variable $\tilde{\tau}$ defined in Sect.~\ref{sec:step1} as regressor variable. For the naive fit, we train a GPR model on the log-scaled luminosity training data, and use it to predict the test data. For the two-step approach, we first train a KNN model to predict the normalized age proxy, $s,$ based on the training data. Second, we use the age proxy prediction as regressor variable when training another GPR model to predict the log-scaled luminosity training data. To compare the outcomes, we plot the prediction of the naive fit and the one resulting from the two-step pipeline separately for each evolutionary phase MS, RGB, and CHeB. For better discrimination of the test data stellar track, neighboring test data points are connected by piecewise linear dashed lines over each phase. The MS evolution data is accurately predicted by both methods, and so is the sub-giant and early red giant phase. The naive fit loses out to the two-step fit during the later stages: the ascension of the RGB and throughout the CHeB phase. \\ Fig.~\ref{fig:core_density_temp} shows that HNNI, by virtue of the same algorithmic prescription, yields accurate forecasts of not only the surface variables $\log L$, $\log T_\mathrm{eff}$, and $\log g$, which have been evaluated in the main part of the paper, but also of all other stellar variables we tested. The shape of the tracks in the $(\log \rho_c, \log T_c)$ diagram is represented well by HNNI, except at fast-timescale transitions during the helium flashes of low-mass stars.

\section{HNNI in three dimensions or higher}
\label{appendix:hnni3d}

Here, we assume that the stellar parameter space is spanned by a third dimension: the initial metallicity $Z_\mathrm{ini}$. The resulting set of parameters $(s, M_\mathrm{ini}, Z_\mathrm{ini})$ allows for a hierarchical ordering: Always first is the age proxy axis, $s$. Always second is the initial mass axis $M_\mathrm{ini}$. Third and least significant out of the three, is the initial metallicity axis $Z_\mathrm{ini}$, whose effect on the shape of stellar evolutionary tracks (at a fixed initial mass) results in minor corrections. The pseudo-code below provides the recipe for a numerical response to the question of what the value of the target variable $Y_j=Y(s_j)$ is at the test location $(s_j, M_\mathrm{ini}^\mathrm{test}, Z_\mathrm{ini}^\mathrm{test})$ in parameter space. Here, $Y$ is any evolved stellar variable: for example, $Y = \log L / \mathrm{L_\odot}$. For full generality, we assume that neither $M_\mathrm{ini}^\mathrm{test}$ nor $Z_\mathrm{ini}^\mathrm{test}$ is contained in the catalog grid database spanned by $\{M_\mathrm{ini}, Z_\mathrm{ini}\}_\mathrm{cat}$. Below, the linear interpolation equation, $y(x) = y_1 + \frac{y_2-y_1}{x_2 -x_1}(x-x_1)$, is referred to by its parameters: $y(x) \leftarrow y_2, y_1, x_2, x_1$. We assume that in the catalog, a similar initial mass grid sampling density is available for each initial metallicity. 
\\
\\
\paragraph{Pseudo-code:}

\begin{enumerate}
    \item Determine the nearest neighbors $Z_\mathrm{ini}^+, Z_\mathrm{ini}^- \in \{ Z_\mathrm{ini} \}_\mathrm{cat}$ to the test initial metallicity value $Z_\mathrm{ini}^\mathrm{test}$ from above and from below, respectively.
    \item From the set of available initial masses $\{ M_\mathrm{ini} \}(Z_\mathrm{ini}^+)$ and $\{ M_\mathrm{ini} \}(Z_\mathrm{ini}^-)$ contained in the catalog at these two metallicities, determine the nearest neighbors to $M_\mathrm{ini}^\mathrm{test}$ from above and from below, respectively:\\
    $M_{+}^+, M_{+}^- \in \{ M_\mathrm{ini} \}(Z_\mathrm{ini}^+)$,\\
    $M_{-}^+, M_{-}^- \in \{ M_\mathrm{ini} \}(Z_\mathrm{ini}^-)$.
    \item From the age proxy series available in the catalog at each of these four initial mass grid points, determine the nearest neighbors to $s_j$ along the age proxy axis that satisfy:\\
    $s_\mathrm{+,max}^+, s_\mathrm{+, min}^+ \in \{ s_i \}(M_{+}^+)$ with $s_\mathrm{+,min}^+<s_j<s_\mathrm{+,max}^+$,\\
    $s_\mathrm{+,max}^-, s_\mathrm{+, min}^- \in \{ s_i \}(M_{+}^-)$ with $s_\mathrm{+,min}^-<s_j<s_\mathrm{+,max}^-$,\\ 
    $s_\mathrm{-,max}^+, s_\mathrm{-, min}^+ \in \{ s_i \}(M_{-}^+)$ with $s_\mathrm{-,min}^+<s_j<s_\mathrm{-,max}^+$,\\
    $s_\mathrm{-,max}^-, s_\mathrm{-, min}^- \in \{ s_i \}(M_{-}^-)$ with $s_\mathrm{-,min}^-<s_j<s_\mathrm{-,max}^-$.
    \item Interpolate along the age proxy axis to find\\ 
    $Y_+^+(s_j) \leftarrow Y_+^+(s_\mathrm{+,max}^+), Y_+^+(s_\mathrm{+,min}^+), s_\mathrm{+,max}^+, s_\mathrm{+,min}^+$,\\
    $Y_+^-(s_j) \leftarrow Y_+^-(s_\mathrm{+,max}^-), Y_+^-(s_\mathrm{+,min}^-), s_\mathrm{+,max}^-, s_\mathrm{+,min}^-$,\\
    $Y_-^+(s_j) \leftarrow Y_-^+(s_\mathrm{-,max}^+), Y_-^+(s_\mathrm{-,min}^+), s_\mathrm{-,max}^+, s_\mathrm{-,min}^+$,\\
    $Y_-^-(s_j) \leftarrow Y_-^-(s_\mathrm{-,max}^-), Y_-^-(s_\mathrm{-,min}^-), s_\mathrm{-,max}^-, s_\mathrm{-,min}^-$.
    \item Interpolate along the log-scaled initial mass axis to find\\
    $Y_+(s_j) \leftarrow Y_+^+(s_j), Y_+^-(s_j), \log (M_{+}^+ / \mathrm{M_\odot}), \log (M_{+}^-/ \mathrm{M_\odot})$,\\
    $Y_-(s_j) \leftarrow Y_-^+(s_j), Y_-^-(s_j), \log (M_{-}^+ / \mathrm{M_\odot}), \log (M_{-}^- / \mathrm{M_\odot})$.
    \item Interpolate along the log-scaled initial metallicity axis to find\\ 
     $Y(s_j) \leftarrow Y_+(s_j), Y_-(s_j), \log (Z_\mathrm{ini}^+/ \mathrm{Z_\odot}), \log (Z_\mathrm{ini}^- / \mathrm{Z_\odot})$.
\end{enumerate}

\paragraph{Generalization:}
The HNNI method is extended analogously to higher-dimensional parameter spaces. We have the number, $n,$ of hierarchical 1D interpolations to perform and number, $k,$ of neighboring grid points to query scale as follows with dimensionality of the parameter space:
\begin{itemize}
    \item 1D $(s)$: $n = 1, k=2,$
    \item 2D $(s, M_\mathrm{ini})$: $n = (1+1)+1=3, k = (2+2)+2=6,$
    \item 3D $(s, M_\mathrm{ini}, Z_\mathrm{ini})$: $n = (3+3)+1 = 7, k = (6 + 6) + 2=14,$  
    \item 4D $(s, M_\mathrm{ini}, Z_\mathrm{ini}, v_\mathrm{ini})$: $n = (7+7)+1 = 15, k=(14 + 14)+2 = 30$.
\end{itemize}
We believe that our HNNI method is generalizable to even higher parameter space dimensions but have not verified this hypothesis. For a binary system composed of two non-rotating stars of the same initial metallicity, we expect the following hierarchical ordering of variables to yield accurate interpolation results:
\begin{equation*}
    \text{5D:} \, (s_1, M_\mathrm{ini,1}, M_\mathrm{ini,2}, P_\mathrm{ini}, \epsilon), 
\end{equation*}
with $n = (15+15)+1=31$, and $k = (30+30)+2=62$.

\section{GPR}
\label{appendix:GPR}

Ever since pioneering work by \cite{Sacks1989}, GPR has been considered standard method choice for emulation tasks because of flexibility of the fitting model and regulatory effect of the Gaussian assumption \citep[for a detailed discussion of application of GPR, see][]{Kennedy2001}. In general, GPR  becomes increasingly time-prohibitive and computationally expensive as the size of training data grows. Particularly, GPR involves the Cholesky factorization and inversion of the covariance matrix, which are computationally costly operations for a large data set.

The literature on GPR includes multiple approaches to improve scalability \citep{8951257}, which can be broadly classified into global approximation and local approximation of the GPR. While approaches to global approximation tend to focus on methods based on sparse kernels \citep{kaufman2008covariance} and approximate likelihoods \citep{varin2011overview}, approaches to local approximation center around inference and prediction on local subsets of data, such as moving-window GPR \citep{van1993site,ni2012moving}. 

Following the local approximation approach, \cite{Li2022} solved the forward problem of stellar evolutionary track forecasts for given fundamental input parameters with separate GPR models that each cover a subspace of the narrow but five dimensional stellar parameter space. However, using a separate GPR on subspaces is likely to ignore the potential dependence across them, which in turn can lead to suboptimal predictions. We expect this to become problematic upon extension of the input space, when exploring parameter spaces of binary star systems. For these reasons, we investigate machine learning models that can be trained on the full data set more closely.

\section{ffNN hyperparameter tuning}
\label{appendix:hyp_tun}

\paragraph{Architecture design:}
There are a number of relevant theoretical considerations that guided our approach to ffNN architecture design. The main role of the activation function is to introduce non-linearities into the information processing pipeline of the neural network. We adopt  the standard recommendation of choosing the ReLU activation function, and instead focus on tuning the model capacity.\footnote{The model capacity corresponds to the number of free trainable parameters. In a fully connected ffNN without regularization layers, the model capacity is given by the total number of weighted connections between neural nodes plus the total number of biases in the network.} There are approaches to tuning the model capacity based on complexity of the learning task \citep{Achille2021}, which can be estimated using the Kolmogorov complexity measure \citep{Kolmogorov1963}. However, its estimation is more a theoretical, less a practical enterprise, due to its non-trivial computation. Instead, there is a body of theoretical hint suggesting that over-parametrization of the deep learning model is required in order to overcome an inherent bias of learning simple (rather than complex) input-output mapping rules \citep{Dingle2018, Nichani2020}. The model capacity ought to be chosen large enough to prevent underfitting, however not overly large to avoid overfitting. The model capacity, once fixed, can be built up in two contrasting ways: 1) by few hidden layers and many neurons per hidden layer or 2) by many hidden layers and few neurons per hidden layer. An incentive toward the first approach is the success of GPR as emulation method: 
a first-order Taylor approximation to the output of a wide network, initialized with independent and identically distributed weights and trained for a large number of epochs, approximates the predictions of a GPR model, and the selection of activation functions corresponds to a particular kernel (the neural tangent kernel) in GPR \citep{Jacot2018, Lee2019}.
The other approach to building up a fixed model capacity, by choosing a higher number of layers,  can, on the other hand, be more valuable than increasing the width. For example, \cite{Eldan2016} show that approximating certain functions requires an exponentially higher number of neurons in a wide network configuration to achieve the same accuracy as that of a deeper network, and the result holds irrespective of the choice of activation function.
We tested both approaches on our problem and found the best result by building up model capacity through a many-layer architecture with a moderate number of neurons. When training deep learning models, we tested dropout, batch normalization, and layer normalization as regularization techniques, in order to push the validation loss further down past stagnation phases. 
\paragraph{Selection criteria:}
We performed empirical tests of manually designed hyperparameter (HP) combinations and then applied selection criteria to decide whether or not to train the configured model up to the end. The main HP that we varied were the number of hidden layers, the number of neurons per layer (assuming a symmetric network architecture), learning rate schedule parameters and the batch size. For each HP combination, we evaluated the loss curve decline during the runtime of learning, and applied the following selection criteria at the $\{500, 10^3, 2 \cdot 10^3, 5 \cdot 10^3 \}$ epoch stages:
First, the speed of learning (judged upon by cross-comparison of validation loss scores at the aforementioned epoch stages for different HP combinations, and estimation of the slopes). Second, the degree of overfit (judged upon by visual assessment of departure of the validation from the training loss curve with increasing epoch number). Subsequently, we manually adjusted the HP choice for the next series of empirical tests, informed by performance of HP combinations from the previous trials. 
Promising models (with fast validation loss curve decline and tolerable overfit over long training periods) were trained until the validation loss curve either flattened out or started to oscillate over epoch scales of order $5-10k$. Out of those promising models, the "best" deep learning model was selected as the one that had the least error scores on the validation data. This procedure was iterated until we trained a surrogate model that attained a threshold value of the validation loss, with the lowest error scores among all the deep learning models we tested over a series of generations. 
Training our best-fit ffNN model lasted around 8 hours on a Nvidia RTX 3060 GPU machine.
\paragraph{Randomness and reproducibility:}
A trained deep learning model is the outcome of a stochastic computer experiment. In order to obtain reproducible results, the random seed needs to be fixed twice: first, before the train-test split of the total data set $N_\mathrm{total} = N_\mathrm{train}+ N_\mathrm{val}$ and second, before initializing the ffNN kernel at model compilation.

\section{Alignment problem}
\label{appenxid:AP}

\paragraph{Globally defined loss functions:}
What remains an issue when building stellar evolution surrogate models with supervised machine learning models to approach the regression problem we formulated, is what we refer to as the alignment problem (AP): our expectation of the surrogate model's predictive capability (characterized by locally accurate performance over all three target variables, across all three evolutionary phases and across the entire initial mass range) does not align with the formalized numerical condition (characterized by the minimization of one single global error score) that is optimized during training of machine learning models. We find that none of the standard loss function choices optimally match our problem setting.
\footnote{
MSE is the average squared residual, where the squared penalization incentivizes to avoid large absolute residuals in model training. Clearly, this behavior is globally desirable for stellar evolutionary track fitting, but leads to too much leniency when a surface variable does not vary much over a star's lifetime. Then, residuals would be small compared to the global range, but large as perceived in HR or Kiel diagrams for a given initial mass. MAE is the average absolute residual, where penalization is linear, and the behavior is reversed in comparison to the MSE. Common scale-free measures, such as MSLE and MAPE, essentially penalize multiplicative errors. MSLE evaluates squared penalties on a log scale (that is, squared log ratios), and MAPE is the average ratio of the absolute residual over the actual value of predicted target variable. In a nutshell, both of these measures tolerate larger absolute residuals as the observed value increases, but we require the opposite for luminosity, which tends stay in a smaller range for tracks at overall high values of luminosity (see top-left panel of Fig.~\ref{fig:err_prop}).
}
The reason for the AP is that the evolution of stars, traced in the HR diagram, does neither happen over the same absolute nor relative  numerical scale range for different initial masses. However, the surrogate model learns by minimizing a globally defined error score, which means by improving to reproduce the overall global shape of the three-dimensional hypersurface of the target variables over the two-dimensional input parameter space. While training deep learning models, we encountered cases when the statistical MSE scores on training and validation data decreased further (i.e., no overfitting in the statistical learning sense of the term), but our physical performance scores, which are locally defined, worsened. In essence, this means that the surrogate model continued to learn, but not that what we appreciate. It may happen that the emulator will have improved predictive capability globally, as assessed by the global loss score, by substantial gains in predictive accuracy in those parameter space regions where the accuracy was already good enough according to our physical performance metrics -- but at the sacrifice of losing predictive accuracy in other parameter space regions admitted by statistical fluctuations. That latter loss in local predictive accuracy, however, may manifest itself in a decrease of physical performance scores over the target variables, adverse to expectations. Nevertheless, this performance loss is not considered problematic by the surrogate model based on the global error score that insufficiently penalizes the prediction errors in relevant parameter regions of concern.\\ The AP is only partially addressed by choosing a ffNN model class, which minimizes the loss of ---not the global data set in a single step, but of--- a sequence of randomly selected data batches\footnote{This point is best understood by comparison of ffNN optimization to that of another statistical learning algorithm. For instance, a RF model is optimized in a single step: a loss score (such as MSE) is minimized after the complete data set is fed into the RF by the bagging technique that distributes the input data onto the individual decision trees.  A RF forecast is an ensemble forecast from an ensemble of decision trees, each of which receives a random split of data samples. For this subset of training data (which differs from one tree to another), the decision tree finds its own hierarchically conditioned numerical rules during the supervised machine learning. However, during training the minimization of the loss score happens globally, not locally for each subset of the total training data set. In contrast, ffNN minimizes the loss score for each batch (a small, randomly selected chunk of the total training data set) and repeatedly over a large number of iterations.}), by choosing the Huber loss score (which seeks a trade-off between MSE and MAE minimization) and by locally increasing the initial mass parameter space sampling (which, statistically, increases the importance of specific parameter space regions by the increased amount of data for that region). If supervised ML is to be applied in high dimensional parameter space for stellar track fitting, this issue requires adequate resolution.

\paragraph{Solution approaches:}
For future extensions of our surrogate modeling method to wide high-dimensional parameter spaces, we propose the following approaches:
\begin{enumerate}
    \item[(i)] To train a separate ffNN model for each target variable, instead of training a single ffNN model with multiple output.
    \item[(ii)] To segment the initial mass parameter space into parts, and train a different ffNN model on each segment of the initial mass range.
    \item[(iii)] To tailor a loss function (parameterized by input variables, in particular the initial mass) to account for differences in numerical scale range over which stellar variables change across the initial mass range, across evolutionary phases and (in the case of multiple output) across different target variables.
\end{enumerate}

Approaches (i) and (ii) are the most common and straightforward. \cite{Li2022} employed both of them when modeling stars by a set of global GPR models. Here, approach (ii) facilitated a splitting of the total training data set of size ${\sim}300k$ into subsets of size ${\sim}20k$, which is their stated limit of computational feasibility for applying global GPR models on a training subset. With approaches (i) and (ii), the learning task is simplified and that can lead to more accurate individual interpolations.\\
However, all three approaches to solving the AP, which can be employed individually or in concert, do have their drawbacks:  For approaches (i) and (ii), many separate models need to be trained, and the capability to capture dependence structures is impaired. 
Approach (iii) is the only solution that leads to truly multivariate predictions, but is the most difficult to realize. If the training target loss does not account for inter-variable dependence, then any loss-based training lacks the required guidance. The first step in accounting for dependence is accounting for variability and covariance. Already that initial step is a challenge since the range of a single stellar surface variable over a star's lifetime can be drastically different depending on initial mass alone. Possibly, creating a suitable multivariate loss function, tailored to stellar evolution tracks, is similarly complex as the multivariate interpolation task itself.

For future research toward extending stellar evolution emulators to wide high-dimensional parameter spaces, we  believe that a good starting approach is to build a separate deep learning model for each target variable and to segment the initial mass range into parts to train sets of univariate deep learning models on each initial mass segment, in an otherwise high-dimensional parameter space.  

\end{appendix}
\end{document}